\pdfoutput=1
\documentclass[fleqn,usenatbib]{mnras}

\usepackage{newtxtext,newtxmath}

\usepackage[T1]{fontenc}

\DeclareRobustCommand{\VAN}[3]{#2}
\let\VANthebibliography\thebibliography
\def\thebibliography{\DeclareRobustCommand{\VAN}[3]{##3}\VANthebibliography}

\usepackage{graphicx}	
\usepackage{amsmath}	
\usepackage{threeparttable}
\usepackage{overpic}   
\usepackage{amsmath}	
\usepackage{makecell}
\usepackage{multicol}
\usepackage{multirow}

\usepackage{color} 
\definecolor{myred}{RGB}{255,66,56}

\usepackage{ulem}

\title[VLBI Astrometry of HD 199178 \& AR Lac]{VLBI Astrometry of Radio Stars to Link Radio and Optical Celestial Reference Frames. I. HD~199178 \& AR~Lacertae}

\author[W. Chen et al.]
{Wen Chen,$^{1,2,3,5,7}~$
Bo Zhang,$^{4}$\thanks{E-mail: zb@shao.ac.cn}
Jingdong Zhang,$^{2,4}$
Jun Yang,$^{5}$
Shuangjing Xu,$^{6,4}$
Yan Sun,$^{4}$
\newauthor Xiaofeng Mai,$^{2,4}$
Fengchun Shu,$^{4}$
and Min Wang$^{1,3,7}$
\\
$^{1}$Yunnan Observatories, Chinese Academy of Sciences, Kunming 650216, Yunnan, China\\
$^{2}$University of Chinese Academy of Sciences, No.19(A) Yuquan Road, Shijingshan District, Beijing 100049, China\\
$^{3}$Key Laboratory for the Structure and Evolution of Celestial Objects, Chinese Academy of Sciences, Kunming 650216, China\\
$^{4}$Shanghai Astronomical Observatory, Chinese Academy of Sciences, 80 Nandan Road, Shanghai, China\\
$^{5}$Department of Space, Earth and Environment, Chalmers University of Technology, Onsala Space Observatory, SE-439 92 Onsala, Sweden\\
$^{6}$Korea Astronomy and Space Science Institute, 776 Daedeok-daero, Yuseong-gu, Daejeon 34055, Republic of Korea\\
$^{7}$Yunnan Key Laboratory of the Solar physics and Space Science, Kunming 650216, China
}

\date{Accepted 2023 April 20. Received 2023 April 03; in original form 2023 January 20}

\pubyear{2023}

\begin{document}
\label{firstpage}
\pagerange{\pageref{firstpage}--\pageref{lastpage}}
\maketitle

\begin{abstract}
To accurately link the radio and optical Celestial Reference Frames (CRFs) at optical bright end, i.e., with \textit{Gaia} $G$ band magnitude \la 13,  increasing number and improving sky distribution of
radio stars with accurate astrometric parameters from both Very Long Baseline Interferometry (VLBI) and 
\textit{Gaia} measurements are mandatory.
We selected two radio stars HD~199178 and AR~Lacertae as the target for a pilot program for the frame link, using the Very Long Baseline Array (VLBA) at 15~GHz at six epochs spanning about 1 year, to measure their astrometric parameters. The measured parallax of HD~199178 is 8.949 $\pm$ 0.059 mas and the proper motion is $\mu_{\alpha}\cos\delta=26.393 \pm 0.093$, $\mu_\delta=-0.950 \pm 0.083$ mas~yr$^{-1}$, while the parallax of AR~Lac is 23.459 $\pm$ 0.094 mas and the proper motion is $\mu_{\alpha}\cos\delta=-51.906 \pm 0.138$, $\mu_\delta=46.732 \pm 0.131$ mas~yr$^{-1}$. Our VLBI measured
astrometric parameters have accuracies about 4-5 times better than the corresponding historic VLBI measurements and comparable accuracies with those from \textit{Gaia}, 
validating the feasibility of frame link using radio stars.
With the updated astrometric parameters for these two stars, there is a $\sim$ 25\% reduction of the uncertainties on the \textit{Y} axis for both orientation and spin parameters.
\end{abstract}




\section{Introduction}
\label{sec1}

How to accurately establish the link (orientation and spin) between radio and optical reference frames is a key question in fundamental astronomy. 
The latest released International Celestial Reference Frame (ICRF) is ICRF3, which is defined by a set of extragalactic radio sources (quasars) with a coordinate noise floor of 30 $\mu$as \citep{2020A&A...644A.159C}. The second release of \textit{Gaia} data (\textit{Gaia} DR2) contains about 0.57 million of quasars, this defines the new-generation optical CRF of \textit{Gaia} DR2 (GCRF2) with a similar coordinate noise floor as ICRF3 \citep{2018A&A...616A..14G}. An updated version, GCRF3, based on the latest \textit{Gaia} early data release 3 (EDR3) contains about 1.61 million quasars \citep{GaiaEDR3}. Although \textit{Gaia} aims to provide a globally consistent CRF for all types of objects, subtle differences among them depend on the magnitude, color and other factors which produce small shifts of image centroids \citep{2016MNRAS.461.1937M}.

The GCRF3 is aligned to the ICRF3 at J2016 through a subset of 2,269 common quasars between radio and optical CRF for the frame orientation, and the spin fixing of GCRF3 is based on the $Gaia$ astrometric solution with constrains of non rotating relative to 429,249 distant quasars, with uncertainties of 10~$\mu$as and 100~$\mu$as~yr$^{-1}$, respectively \citep{GaiaEDR3}. In order to realize a consistent CRF at different magnitudes in \textit{Gaia} EDR3, an ad hoc correction for the Window Class (WC) effects of about 100~$\mu$as~yr$^{-1}$ was performed \citep{2020A&A...633A...1L}. The correction was estimated from a comparison of the stellar proper motions between \textit{Gaia} EDR3 and Hipparcos, with a spin uncertainty of about 24~$\mu$as~yr$^{-1}$ for the bright GCRF3. However, this spin uncertainty can not be decreased, since it is dominated by the large uncertainties of the Hipparcos frame, hence the WC correction method will be of limited value for the validation of the ﬁnal GCRF. In order to achieve a stellar frame link accuracy comparable to that of quasars, the uncertainties of spin parameters should be <10 $\mu$as yr$^{-1}$. This highlights the requirement for independent methods, e.g.,
Very Long Baseline Interferometry (VLBI), which has the comparable or even better astrometric accuracy with \textit{Gaia}, to verify the consistency of the GCRF~\citep{2020A&A...633A...1L}.


As a pilot program for VLBI astrometry of radio stars to 
link the optical and radio stellar CRFs, we observed HD~199178 and AR~Lacertae (hereafter AR~Lac) using the National Radio Astronomy Observatory (NRAO)'s\footnote{The National Radio Astronomy Observatory is a facility of the National Science Foundation (NSF) operated under cooperative agreement by Associated Universities, Inc.} Very Long Baseline Array (VLBA), to study the optimum observing strategy and demonstrate the accuracy that can be achieved. 
The two radio stars play an important role which contribute the most weight to the spin solution in \cite{2020A&A...633A...1L} to link (orientation and spin) between radio and optical reference frames. However, both uncertainties of parallaxes and proper motions
from previous VLBI observations by~\citet{1999A&A...344.1014L}   
are about one order of magnitude larger than those of \textit{Gaia} DR3, precluding an
accurate link of the stellar CRFs. 

HD~199178 (V1794~Cyg) is one of the FK Comae-type stars, which form the very small group consisting of stars that are single, rapidly rotating, and extremely active G–K-type subgiants or giants, it has an active magnetic strong emission from the corona and transition region \citep{2019A&A...625A..79H}.

AR Lacertae (HIP 109303) is an eclipsing close binary (G2IV + K0IV) of the RS~CVn type with a period of $\sim$1.98 days and an edge-on approximate circular orbit with a semi-major axis of $\sim$8.87$R_{\odot}$ \citep{2006BaltA..15..387S} and an inclination of 89\fdg4~\citep{2008MNRAS.389.1722E}. Its binary orbit period can be characterized by 
a linear decreasing trend plus a sinusoidal function with period of decades~\citep{2006BaltA..15..387S,2012PASJ...64...84L}, and there might be a third companion
\citep{2020JPhSt..24.3904T}, i.e., a white dwarf  21\farcs 9 away with a orbit period of 16,000~yrs~\citep{2022RNAAS...6..127B}.

In this paper, we report the updated astrometric parameters of HD~199178 and AR~Lac
from VLBA observations at six epochs. In section \ref{sec:obs}, we show the strategy and setup of the VLBA observations, and then describe the steps of data reduction. In section \ref{sec:result}, we present the images of the radio stars and their background calibrators, and then derive their proper motions and parallaxes from the time series of the stellar position offset to the background sources. In section \ref{sec:disc}, we discuss the possible astrometric error sources. Finally, we summarize our results in section \ref{sec:sum}.

\section{Observations and data reduction}
\label{sec:obs}

\subsection{VLBA observations at 15 GHz}
\label{ssec:obs_vlba} 

The VLBI observations were carried out under the VLBA program BZ077
at 15~GHz,  which is suitable for balancing the array sensitivity and angular resolution
for observing the continuum emission from the radio stars. 
To better sample the data for parallax fit but not request too much observing time, we
observe the two stars in a common observation at six epochs, which are 
close to the peaks of the parallax curve in Right Ascension (RA). This allows increasing the
sensitivity of parallax measurement, since for most cases, the astrometric accuracy in
RA is much better than in Declination (Dec), mainly owing to the shape of synthesis beam, i.e., a better angular resolution in RA for VLBA with the longest baselines in East-West directions. 
Table~\ref{Ants_codes} lists the information of the six VLBA observations. 
Each VLBI observation involves 8-10 VLBA telescopes, with a recording rate of 2048 Mbps (16 sub-bands in dual polarization, 32 MHz per sub-band, and 2 bits quantization).

In order to accurately measure the positions of the target stars with respect to the background calibrators, the observations were performed in the phase-referencing mode~\citep{Beasley1994}, i.e.,
for our case, by rapidly nodding observations between a pair of nearby sources with a cycle time of about 2 minutes, to determine phase solutions
from the calibrator and then apply them to the stars. To reduce systematic positional errors which are dependent on source separation, we selected two calibrators for each target, i.e., J2040$+$4527 and J2102$+$4702 for HD~199178, and J2202$+$4216 and J2153$+$4322 for AR~Lac. 
These calibrators have angular separations of 2\fdg5 -- 3\fdg6 from the targets.
Table~\ref{calibs_coords} lists the ICRF positions and their uncertainties of these calibrators.

\begin{figure*}
\centering
 \includegraphics[width=\linewidth]{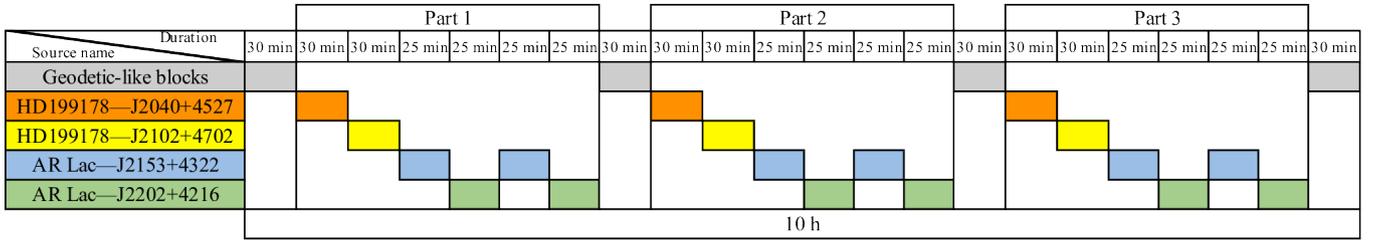}
 \caption{The diagram of observing time arrangement for each epoch. Gray blocks represent the geodetic-like blocks, orange blocks represent the scans of the radio star HD199178 and its calibrator J2040+4527, yellow blocks represent the scans of HD199178 and another calibrator J2102+4702; blue blocks represent the scans of the radio star AR Lac and its calibrator J2153+4322, and green blocks represent the scans of AR Lac and another calibrator J2202+4216; the total observation time of each epoch is 10 hours. The time on the diagram includes the antennas slewing time.}
 \label{time_sched}
\end{figure*}

Figure~\ref{time_sched} demonstrates the observing strategy at each epoch. We divided the 10-h observations into three 160-min phase-referencing
and four 30-min geodetic-like blocks. In the phase-referencing blocks, we alternately observe
the two radio stars and their background quasars;
while in the geodetic-like blocks, we observe $\sim$10 ICRF quasars whose positions are known to be better than 1 mas, to estimate
the variable wet tropospheric zenith delays, which are dominant error sources
but not taken into account 
in the VLBA correlation model~\citep{2009ApJ...693..397R}.
The typical total on-source time was about 90~min for HD~199178 while
137~min for AR Lac.

\begin{table}
	\centering
	\caption{Dates and participated antennas of the VLBA observations}
	\label{Ants_codes}
	\resizebox{\hsize}{!}{  
        \begin{threeparttable}
	\begin{tabular}{ccc}
	\hline
  	 Epoch  & Date      & Stations \\
        \hline
	A & 2020-05-09          &  \texttt{BR, FD, HN, KP, LA, MK, NL, OV, PT, SC} \\ 
	B & 2020-05-17          &  \texttt{BR, FD, HN, KP, LA, MK, NL, OV, PT, SC} \\ 
	C & 2020-11-05          &  \texttt{---\tnote{*}, FD, HN, KP, LA, MK, NL, ---\tnote{*}, PT, SC} \\ 
	D & 2020-11-09          &  \texttt{BR, FD, HN, KP, LA, MK, NL, ---\tnote{*}, PT, SC} \\ 
	E & 2021-05-09          &  \texttt{BR, FD, HN, KP, LA, MK, NL, OV, PT, SC} \\ 
	F & 2021-06-08          &  \texttt{BR, FD, HN, KP, LA, MK, NL, OV, PT, SC} \\ 
	\hline
	\end{tabular}
	\begin{tablenotes}    
        \footnotesize               
        \item[*] Absent in the observation.          
        \item[~] The two-letters codes for each station: Brewster (\texttt{BR}), Fort Davis (\texttt{FD}), Hancock (\texttt{HN}), Kitt Peak (\texttt{KP}), Los Alamos (\texttt{LA}), Mauna Kea (\texttt{MK}), North Liberty (\texttt{NL}), Owens Valley (\texttt{OV}), Pie Town (\texttt{PT}) and Saint Croix (\texttt{SC})
    \end{tablenotes}
    \end{threeparttable}
    }  
\end{table}

\begin{table*}
	\centering
	\caption{Summary of the list of the calibrators}
	\label{calibs_coords}
	\small
	\begin{threeparttable}
	\begin{tabular}{ccccccc}
	\hline
  	  Radio & Calibrator & $\alpha$\  & $\sigma_{\alpha}$  & $\delta$     & $\sigma_{\delta}$ & Separation \\
        Stars &             & (h m sec)    &(mas)        &($^{\circ}$ ' ")   & (mas)         &(degree)  \\
    \hline
    HD199178 & J2040+4527 & 20 40 48.3331 & & 45 27 17.147                &  & 2.55 \\
             & J2102+4702 & 21 02 17.0560 &  & 47 02 16.254               &  & 3.03 \\
    \hline
     AR Lac  & J2153+4322 & 21 53 50.9591 &  & 43 22 54.501                &     & 3.56 \\ 
	       & J2202+4216 & 22 02 43.2914 & & 42 16 39.980                &    & 3.63 \\
    \hline
    HD199178 & J2040+4527 & 20 40 48.333108  &0.18 & 45 27 17.14587    &0.13  & 2.55 \\
             & J2102+4702 & 21 02 17.056003 &0.16  & 47 02 16.25324    &0.11  & 3.03 \\
    \hline
     AR Lac  & J2153+4322 & 21 53 50.959117  &0.16  & 43 22 54.50085   &0.12    & 3.56 \\ 
	       & J2202+4216 & 22 02 43.291372 &0.14 & 42 16 39.97993    &0.10    & 3.63 \\
	\hline
	\end{tabular}
	\begin{tablenotes}    
        \footnotesize               
        \item \textbf{NOTE.} The first four lines of the coordinates of calibrators we used in the VEX files during the VLBI observations, while the last four lines of the updated coordinates and errors we used in frame link are from rfc\_2022d ( \url{http://astrogeo.org}).
    \end{tablenotes}
	\end{threeparttable}
\end{table*}

\subsection{Data correlation and reduction}
\label{ssec:corr} 

The raw data were correlated by the VLBA DiFX software correlator \citep{2007PASP..119..318D} in Socorro, NM, which generated 64 channels in total with 32~MHz in each of 16 (dual-polarisation) IFs. 
We calibrate the visibility data using the NRAO software package Astronomical Image Processing System \citep[AIPS,][]{Greisen2003} and make images using \textsc{difmap} \citep[version 2.5e, ][]{1994BAAS...26..987S}.   

To calibrate the visibility amplitude calibration, we adopted the calibration tables with 
system temperature and antenna gain curve information, to convert correlation coefficients to flux density units. For phase calibration, we first reduced the geodetic-like blocks data to estimate the residual zenith atmospheric delays according to the procedure as described by \citep{2009ApJ...693..397R} in four steps. In step 1, we correct the ionospheric dispersive delays using the map of total electron content provided by the Global Positioning System (GPS) satellite observations, and then correct the phases due to antenna parallactic angle variations and not accurate
Earth Orientation Parameters. In step 2, we did the ``manual phase calibration'' from a scan on a strong
fringe-finder J2202$+$4216 to remove the instrumental delay and phase differences among IFs. In step 3, we carried out the global fringe fitting and printed out the solution of delays and rates. Finally, Based on the delay and rate data, we used a least-squares program to solve for the clock and zenith atmospheric delays, the results were saved in a text file, which was used for calibrating the atmospheric delays for the data from the phase-reference blocks.

For the phase-reference blocks data, steps 1 and 2 are the same as for the geodetic-like blocks.
In step 3, 
we performed the global fringe-fitting on all calibrators. In step 4, we exported the calibrated data and imaged every calibrator in \textsc{difmap} \citep[version 2.5e, ][]{1994BAAS...26..987S}. In step 5, we loaded the image fits files of calibrators into \textsc{aips} and carried out ``self-calibration'' on calibrators to remove any structure phase errors.  In step 6, we exported the final calibrated data and made images with natural weighting for both calibrators and targets using \textsc{difmap}.

\begin{table*}
\caption{Positions offsets and flux densities of AR~Lac and HD~199178.}
    \begin{threeparttable}
	\centering
	\label{tab:targets_coords}
	\small
	\begin{tabular}{cccccccccc}

\hline
Radio   & Background  & Date   & East Offset          & North Offset    & Total Flux  & Peak Intensity    & Image & Model size & P.A. \\
Star    & Source      & (year)  & (mas)                &(mas)            & (mJy)          & (mJy~beam$^{-1}$)  & SNR &  (mas)     & (degree)\\
\hline
HD199178 & J2040+4527 & 2020.35331 & -3.828 $\pm$ 0.008    &  2.793 $\pm$ 0.009 & 16.09 $\pm$ 0.82  &9.08 & 41.48 &1.3$\times$0.9 &151.4  \\
         &            & 2020.37510 & -3.813 $\pm$ 0.015    &  4.172 $\pm$ 0.017 & 6.70 $\pm$ 0.37  &3.31 & 25.12 &1.4$\times$0.9 &146.4  \\
         &            & 2020.84746 & -8.020 $\pm$ 0.034    & -1.358 $\pm$ 0.035 & 2.54 $\pm$ 0.24  &1.48 & 12.68 &1.1$\times$1.1 &2.5  \\
         &            & 2020.85843 & -7.961 $\pm$ 0.017    & -1.702 $\pm$ 0.018 & 8.69 $\pm$ 0.49  &4.79 & 25.85 &1.4$\times$0.9 &139.9  \\
         &            & 2021.35261 & 22.399 $\pm$ 0.017    &  1.968 $\pm$ 0.024 & 2.14 $\pm$ 0.15  &1.62 & 20.26 &1.3$\times$0.8 &157.9  \\
         &            & 2021.43443 & 22.946 $\pm$ 0.015    &  5.459 $\pm$ 0.020 & 2.63 $\pm$ 0.18  &1.68 & 20.25 &1.2$\times$0.9 &174.1  \\
\hline 
HD199178 & J2102+4702 & 2020.35338 & -3.652 $\pm$ 0.006    &  2.524 $\pm$ 0.007 & 14.19 $\pm$ 0.72  &10.30 & 43.39 &1.0$\times$0.8 &171.4  \\   
         &            & 2020.37516 & -3.718 $\pm$ 0.011    &  3.887 $\pm$ 0.014 & 6.83 $\pm$ 0.37  &3.77 & 24.32 &1.2$\times$0.9 &164.0  \\
         &            & 2020.84752 & -7.742 $\pm$ 0.025    & -1.686 $\pm$ 0.025 & 2.03 $\pm$ 0.16  &1.37 & 18.59 &1.1$\times$0.8 &135.4  \\
         &            & 2020.85850 & -7.828 $\pm$ 0.008    & -1.979 $\pm$ 0.010 & 8.44 $\pm$ 0.47  &4.56 & 29.46 &1.1$\times$0.8 &153.1  \\
         &            & 2021.35268 & 22.544 $\pm$ 0.016    &  1.778 $\pm$ 0.021 & 1.80 $\pm$ 0.13  &1.42 & 19.44 &1.1$\times$0.8 &171.9  \\
         &            & 2021.43449 & 23.074 $\pm$ 0.012    &  5.203 $\pm$ 0.019 & 2.37 $\pm$ 0.16  &1.67 & 22.97 &1.2$\times$0.7 &7.0  \\               
\hline 
AR Lac   & J2153+4322 & 2020.35347 & 46.417 $\pm$ 0.009    &   -22.417 $\pm$ 0.011   & 5.34 $\pm$ 0.28   &3.20 & 31.77 &1.1$\times$0.9 &2.6 \\ 
         &            & 2020.37526 & 45.960 $\pm$ 0.009    &   -18.616 $\pm$ 0.012   & 9.31 $\pm$ 0.48  &4.43 & 31.13 &1.2$\times$0.9 &2.9  \\
         &            & 2020.84761 & -21.551 $\pm$ 0.020   &    1.475 $\pm$ 0.025    & 4.70 $\pm$ 0.28   &1.57 & 20.47 &1.5$\times$0.9 &151.8  \\
         &            & 2020.85858 & -23.177 $\pm$ 0.010  &    0.005 $\pm$ 0.012    & 25.41 $\pm$ 1.28  &7.59 & 26.60 &1.3$\times$1.0 &160.5  \\
         &            & 2021.35276 &  -5.234 $\pm$ 0.015   &    24.416 $\pm$ 0.017   & 2.60 $\pm$ 0.16   &1.68 & 23.68 &1.2$\times$0.9 &149.9  \\
         &            & 2021.43458 & -10.118 $\pm$ 0.012   &    37.973 $\pm$ 0.014   & 6.45 $\pm$ 0.34   &2.69 & 29.12 &1.4$\times$1.1 &158.5  \\
\hline          
AR Lac   & J2202+4216 & 2020.35351 & 46.908 $\pm$ 0.007    &   -22.588 $\pm$ 0.010   & 5.09 $\pm$ 0.27   &3.38 & 40.38 &1.1$\times$0.8 &7.8  \\  
         &            & 2020.37530 & 46.451 $\pm$ 0.009    &   -18.834 $\pm$ 0.011   & 8.82 $\pm$ 0.46   &4.67 & 29.47 &1.1$\times$0.9 &175.8  \\
         &            & 2020.84766 & -21.097 $\pm$ 0.015   &    1.191 $\pm$ 0.024    & 4.44 $\pm$ 0.27   &1.79 & 20.49 &1.5$\times$0.8 &162.3  \\
         &            & 2020.85863 & -22.710 $\pm$ 0.009   &   -0.259 $\pm$ 0.010    & 25.58 $\pm$ 1.29  &7.90 & 34.23 &1.3$\times$1.0 &154.3  \\
         &            & 2021.35281 & -4.727 $\pm$ 0.017    &    24.136 $\pm$ 0.020   & 2.73 $\pm$ 0.17   &1.63 & 19.55 &1.3$\times$0.8 &149.2  \\
         &            & 2021.43463 & -9.602 $\pm$ 0.010    &    37.723 $\pm$ 0.012   & 7.20 $\pm$ 0.38   &2.73 & 37.08 &1.4$\times$1.1 &158.9  \\          
	\hline
	\end{tabular}
		\begin{tablenotes}    
        \footnotesize               
        \item \textbf{Note.} The fourth and fifth columns give position offsets relative to $\alpha$ = $20^{\rm h}53^{\rm m}53\fs7034$, $\delta$ = $+44\degr23\arcmin11\farcs0630$ for HD199178, and $\alpha$ = $22^{\rm h}08^{\rm m}40\fs7141$, $\delta$ = $+45\degr44\arcmin33\farcs0870$ for AR Lac. The ninth and tenth columns give the angular size of the Gaussian model and the position angle from \texttt{JMFIT}. The quoted offset uncertainties are only formal errors determined from the phase-referenced images. The uncertainties of the total flux densities included formal errors and 5\% systematic errors.
      \end{tablenotes}
     \end{threeparttable}
\end{table*}

\begin{table*}
	\centering
	\caption{Astrometric parameters of HD199178 and AR Lac.}
	\label{tab:targets_fitted_results}
	\small
	\begin{threeparttable} 
	\begin{tabular}{ccc ccc ccc}
	\hline
  	   Radio  &$\Pi$  & $\mu_x$           & $\mu_y$         &  Systematic                & Background & Observation & Telescope \\
       Star       & (mas)               & (mas~yr$^{-1}$)      & (mas~yr$^{-1}$) &  Errors (mas)              & Source     &  Period        &           &\\
	\hline
	HD199178 & 8.975 $\pm$ 0.079    & 26.395 $\pm$ 0.172   & -0.980 $\pm$ 0.190 & & J2040+4527   & 2020/05-2021/06 &  VLBA~\tnote{1} \\
	HD199178 & 8.927 $\pm$ 0.101    & 26.392 $\pm$ 0.231   & -0.921 $\pm$ 0.174 & & J2102+4702   & 2020/05-2021/06 &  VLBA~\tnote{1} \\
	HD199178 & 8.949 $\pm$ 0.059   & 26.393 $\pm$ 0.093  &  -0.950 $\pm$ 0.083 &  0.191/0.169~\tnote{6} & combined two  & 2020/05-2021/06 & VLBA~\tnote{1} \\
	HD199178 & 8.59~~ $\pm$ 0.33~~     & 26.60~~ $\pm$ 0.41~~   &  -1.24~~ $\pm$ 0.43~~   & 0.35 & J2102+4702  & 1992/09–1994/09 & global VLBI~\tnote{2}\\
	HD199178 &                     & 27.12~~ $\pm$ 6.14~~    &  -1.03~~ $\pm$ 3.51~~   & & J2038+5119  & 2003/06-2004/10 &VLA+PT~\tnote{3} \\
	HD199178 & 8.891 $\pm$ 0.015  & 26.451 $\pm$ 0.015   & -0.877 $\pm$ 0.015 & 0.086~\tnote{7}  &           & 2014/07–2017/05 & \textit{Gaia} DR3\tnote{4}\\
	& & & & & & & & \\
        AR Lac   & 23.448 $\pm$ 0.148   & -51.917 $\pm$ 0.307   & 46.767 $\pm$ 0.298 &                   & J2153+4322 & 2020/05-2021/06  & VLBA~\tnote{1} \\
	AR Lac   & 23.469 $\pm$ 0.145   & -51.895 $\pm$ 0.303   & 46.697 $\pm$ 0.278 &                   & J2202+4216 & 2020/05-2021/06 & VLBA~\tnote{1} \\
	AR Lac   & 23.459 $\pm$ 0.094   & -51.906 $\pm$ 0.138   & 46.732 $\pm$ 0.131 &  0.285/0.267~\tnote{6} & combined two & 2020/05-2021/06 & VLBA~\tnote{1} \\
	AR Lac   & 23.97~~ $\pm$ 0.37~~ & -52.08~~ $\pm$ 0.13~~ & 47.03~~ $\pm$ 0.19~~ &      0.44              &J2202+4216 & 1989/04–1994/05 & global VLBI~\tnote{2}\\
        AR Lac   &                      & -51.13~~ $\pm$ 1.42~~ & 47.36~~ $\pm$ 1.42~~ &                &J2202+4216 & 2001/10-2003/06 & MERLIN~\tnote{5} \\
	AR Lac   &                      & -52.43~~ $\pm$ 1.02~~ & 46.77~~ $\pm$ 1.07~~ &                    &J2202+4216 & 2003/06-2004/10 & VLA+PT~\tnote{3} \\
	AR Lac   & 23.525 $\pm$ 0.023 & -52.310 $\pm$ 0.021  & 46.931 $\pm$ 0.019 & 0.173~\tnote{7} &            & 2014/07–2017/05 &\textit{Gaia} DR3\tnote{4}\\

	\hline
	\end{tabular}
	\begin{tablenotes} 
        \footnotesize 
        \item \textbf{Note.} $\Pi$ represents parallax, absolute proper motions are defined as $\mu_x$=$\mu_{\alpha}\cos\delta$  and $\mu_y$=$\mu_\delta$. The offsets relative to the 
        reference coordinates in Table~\ref{tab:targets_coords} at middle epochs, i.e., 
         $\Delta x=1.69\pm0.08$~mas, $\Delta y=0.25\pm0.07$~mas at 2020.89390 for HD199178 and $\Delta x=-2.99\pm0.12$~mas, $\Delta y=2.57\pm0.11$~mas at 2020.89404 for AR~Lac, are determined from the
        ``combined two'' parallax fits and corrected from frame link.
        \item[1] This paper.          
        \item[2] \cite{1999A&A...344.1014L}.
        \item[3] \cite{VLA+PT_2}.        
        \item[4]\cite{GaiaEDR3}. 
        \item[5] \cite{merlin_astrometry}, combined with \cite{VLA+PT_2} to get proper motion.
        \item[6] Error floors in $\alpha$/$\delta$ (see Section~\ref{ssec:astrometry}).
        \item[7] \textit{Gaia} astrometric excess noise~\citep{2022gdr3.reptE..20H}.
      \end{tablenotes} 
	\end{threeparttable}
\end{table*}

\begin{table*}
	\centering
	\caption{Coordinates of the radio stars derived from \texttt{JMFIT} at the middle epoch 2020-11-09 from the VLBA observation.}
	\label{tab:Coords_radiostars}
	\small
	\begin{threeparttable}
	\begin{tabular}{ccccccc}
	\hline
  	   Radio  & $\alpha$  & $\sigma_{\alpha}$  & $\delta$     & $\sigma_{\delta}$  & Calibrator \\
        star      &  (h m s)    &(mas)          &(\degr~\arcmin~\arcsec)   & (mas)         &   \\
    \hline
     HD199178 &20 53 53.7026574 &0.017 & 44 23 11.061298  &0.018  & J2040+4527 \\
               &20 53 53.7026698 &0.008 & 44 23 11.061021  &0.010   & J2102+4702 \\
    \hline
     AR Lac  & 22 08 40.7118859 &0.010 & 45 44 33.087005  & 0.012  & J2153+4322 \\
              & 22 08 40.7119305 &0.009 & 45 44 33.086742  & 0.010  & J2202+4216 \\
	\hline
	\end{tabular}
	\begin{tablenotes}    
        \footnotesize               
        \item \textbf{Note.} An example for the coordinates of the two radio stars, the coordinates are derived by adding the position offsets on the reference position in Table~\ref{tab:targets_coords}, $\sigma_{\alpha}$ and $\sigma_{\delta}$ are formal errors from \texttt{JMFIT}.          
        \end{tablenotes}
	\end{threeparttable}
\end{table*}

\section{Results}
\label{sec:result}

\begin{figure*}
	\centering
		\centering
		\begin{overpic}[width=0.30\linewidth]{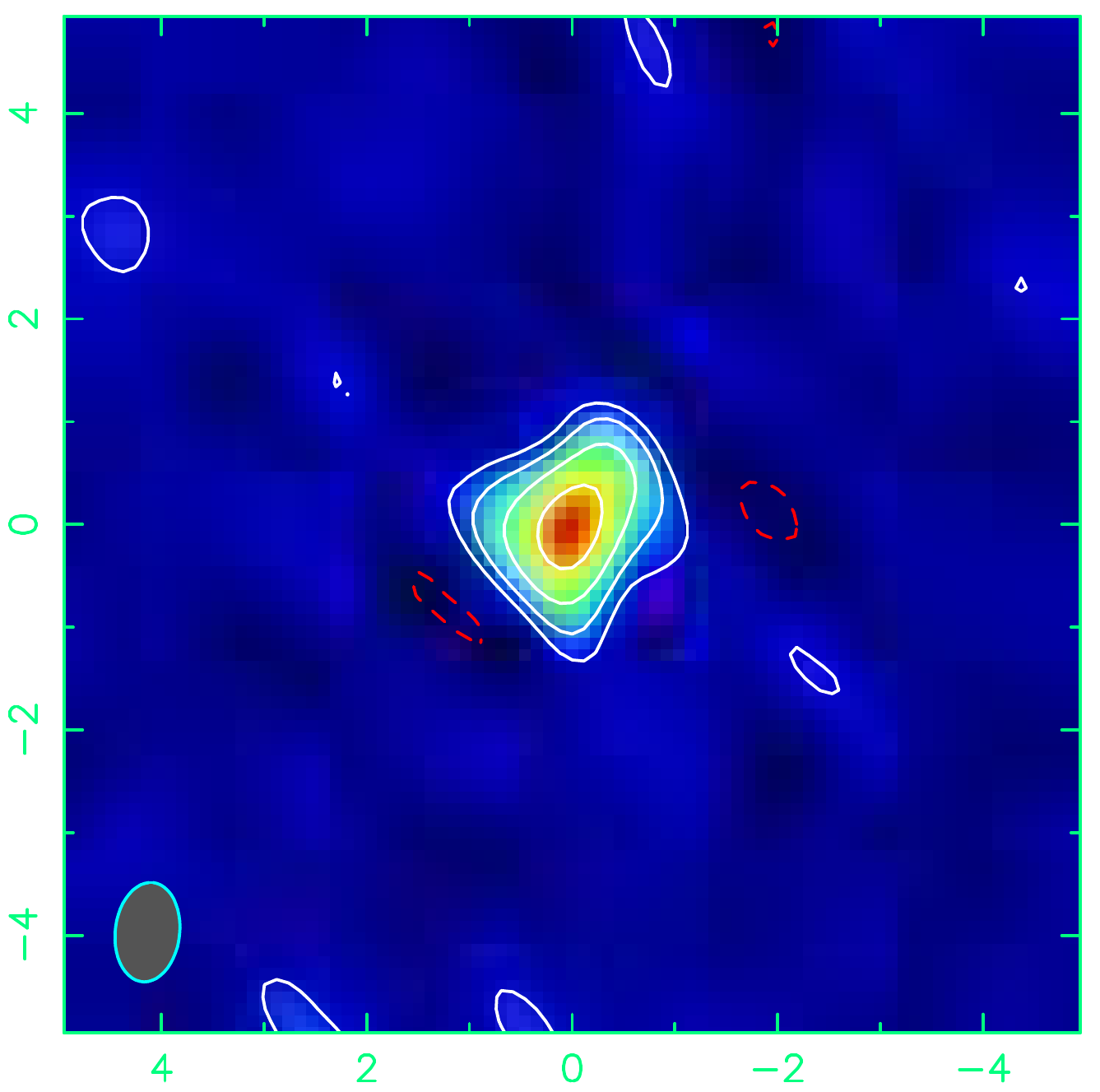}
		\put(12,90){\small\color{white}{HD199178}}
        \put(70,90){\small\color{white}{2020-05-09}}
        \put(45,10){\small\color{white}{Calibrator: J2040+4527}}
        \end{overpic}
        \begin{overpic}[width=0.30\linewidth]{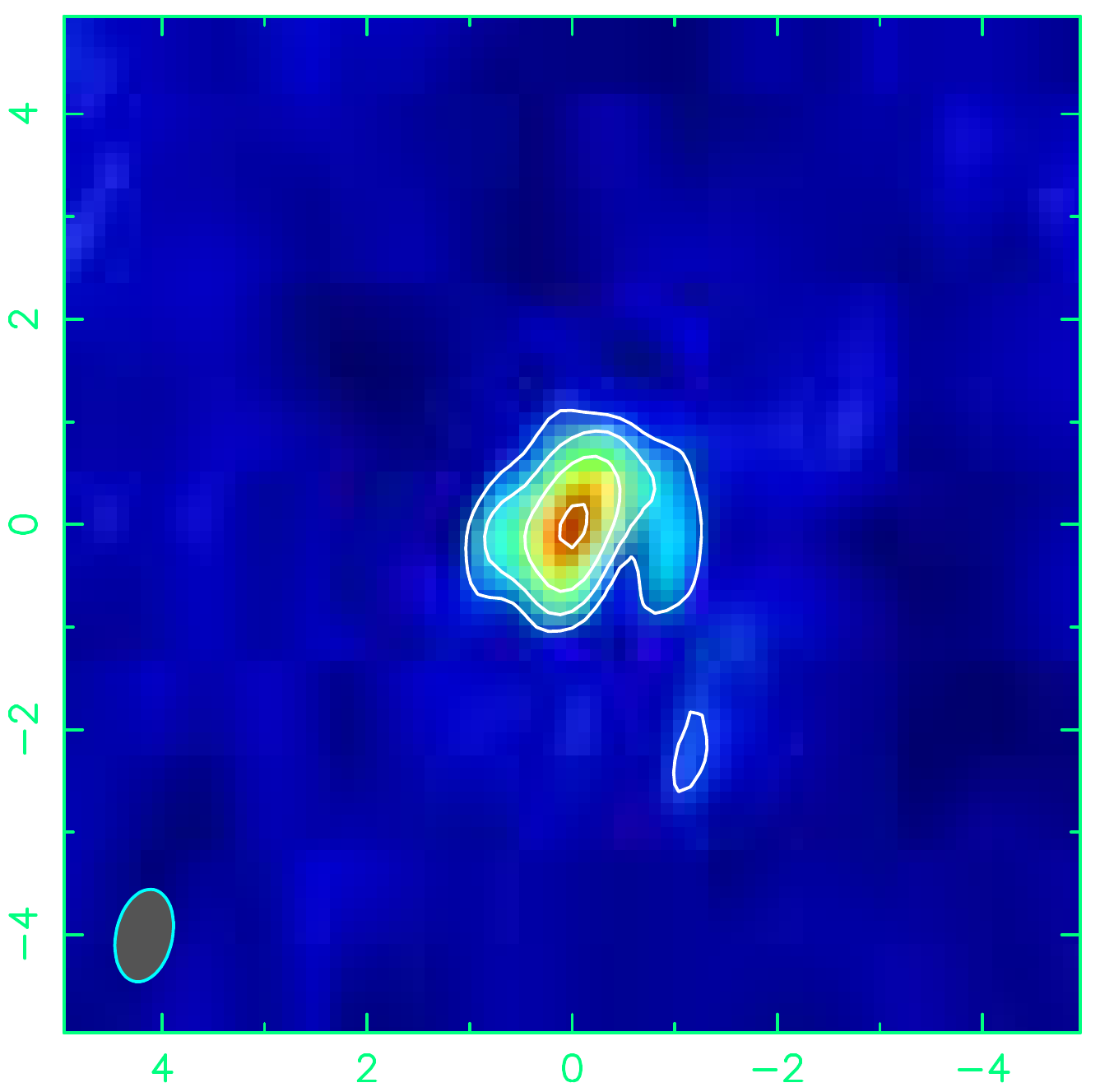}
		\put(12,90){\small\color{white}{HD199178}}
        \put(70,90){\small\color{white}{2020-05-17}}
        \put(45,10){\small\color{white}{Calibrator: J2040+4527}}
        \end{overpic}
        \begin{overpic}[width=0.30\linewidth]{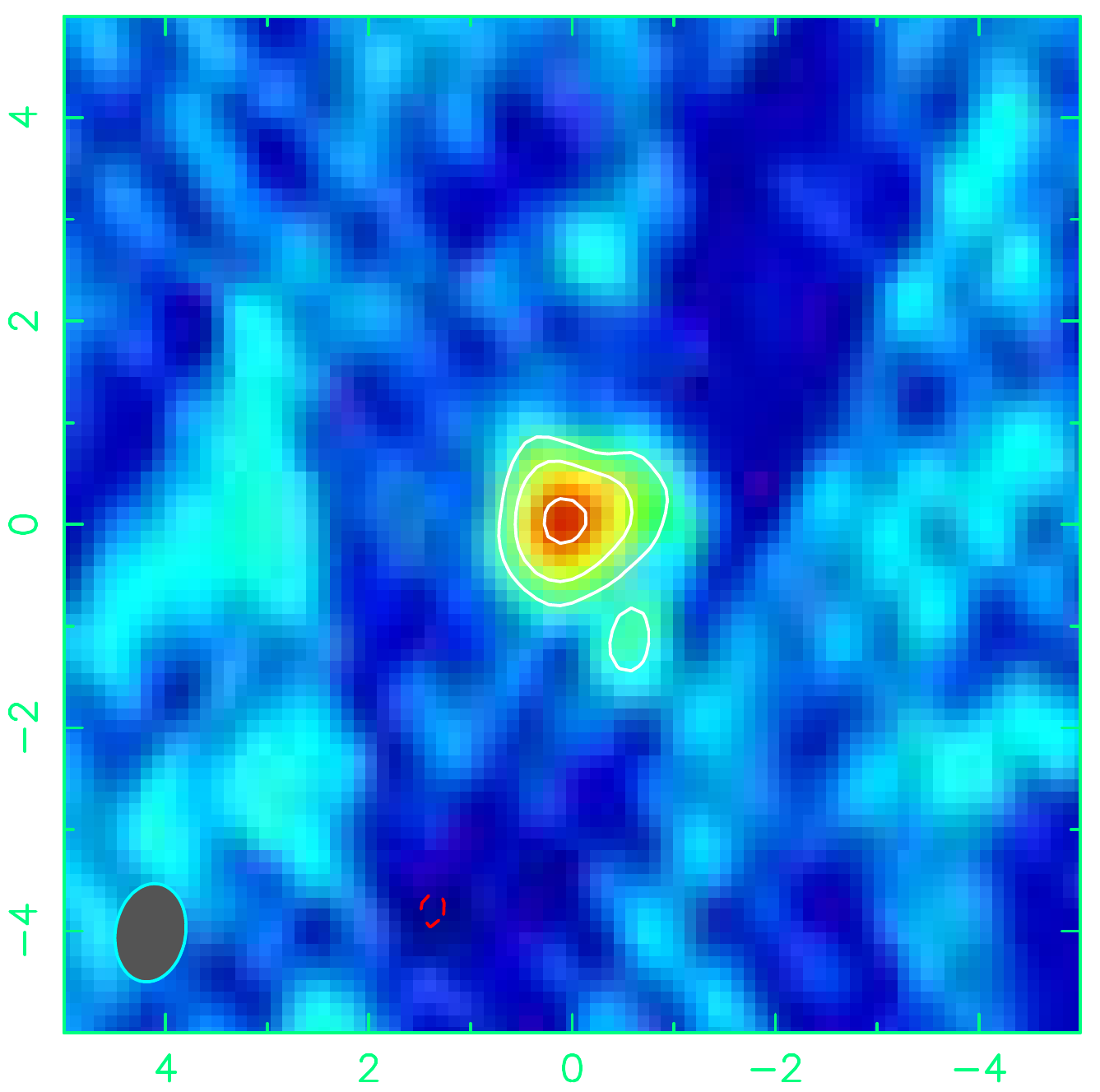}
		\put(12,90){\small\color{white}{HD199178}}
        \put(70,90){\small\color{white}{2020-11-05}}
        \put(45,10){\small\color{white}{Calibrator: J2040+4527}}
        \end{overpic}
		\label{HD_row1}
	
	\centering
		\centering
		\begin{overpic}[width=0.30\linewidth]{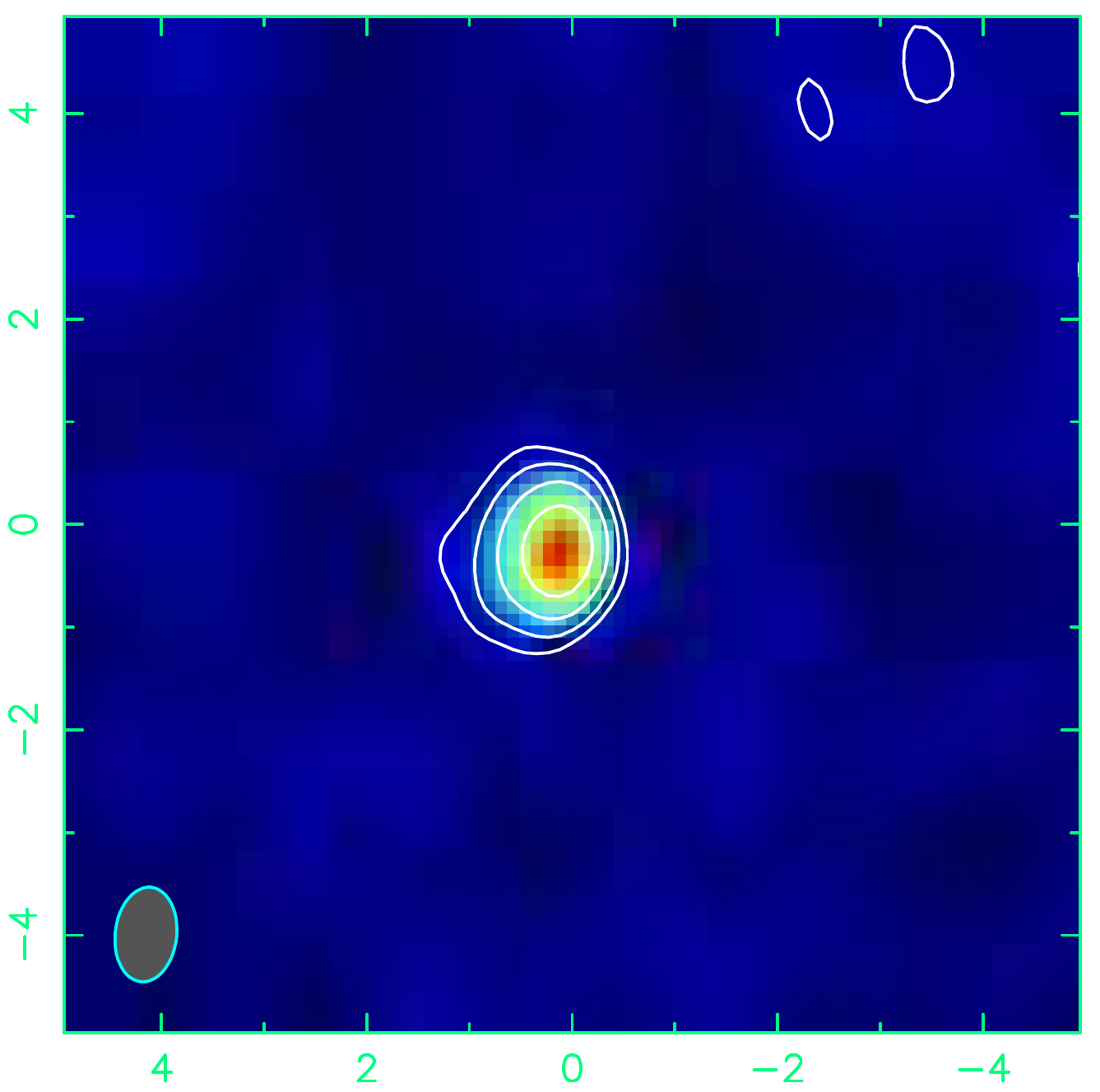}
		\put(12,90){\small\color{white}{HD199178}}
        \put(70,90){\small\color{white}{2020-05-09}}
        \put(45,10){\small\color{white}{Calibrator: J2102+4702}}
        \end{overpic}
        \begin{overpic}[width=0.30\linewidth]{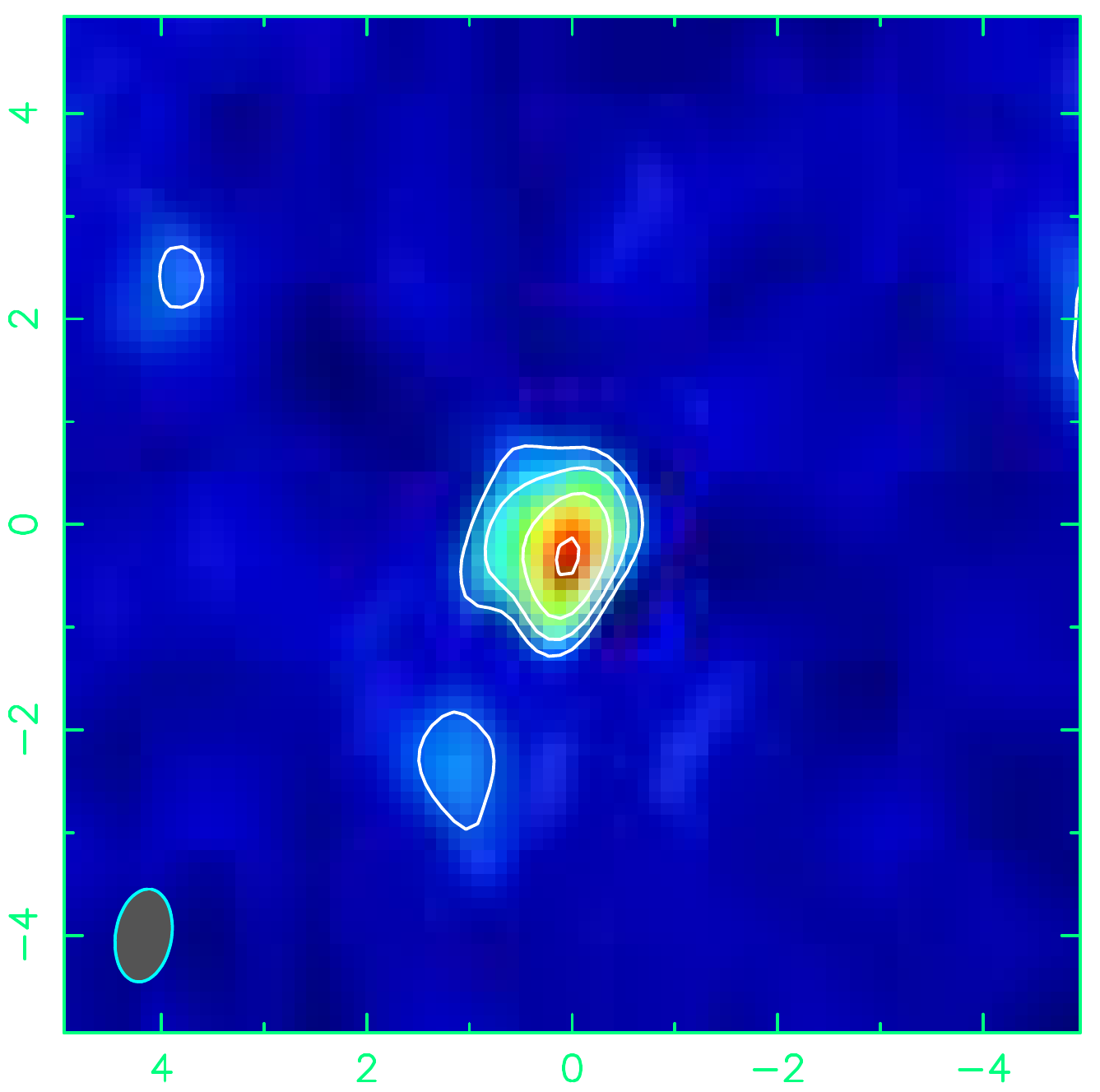}
		\put(12,90){\small\color{white}{HD199178}}
        \put(70,90){\small\color{white}{2020-05-17}}
        \put(45,10){\small\color{white}{Calibrator: J2102+4702}}
        \end{overpic}
        \begin{overpic}[width=0.30\linewidth]{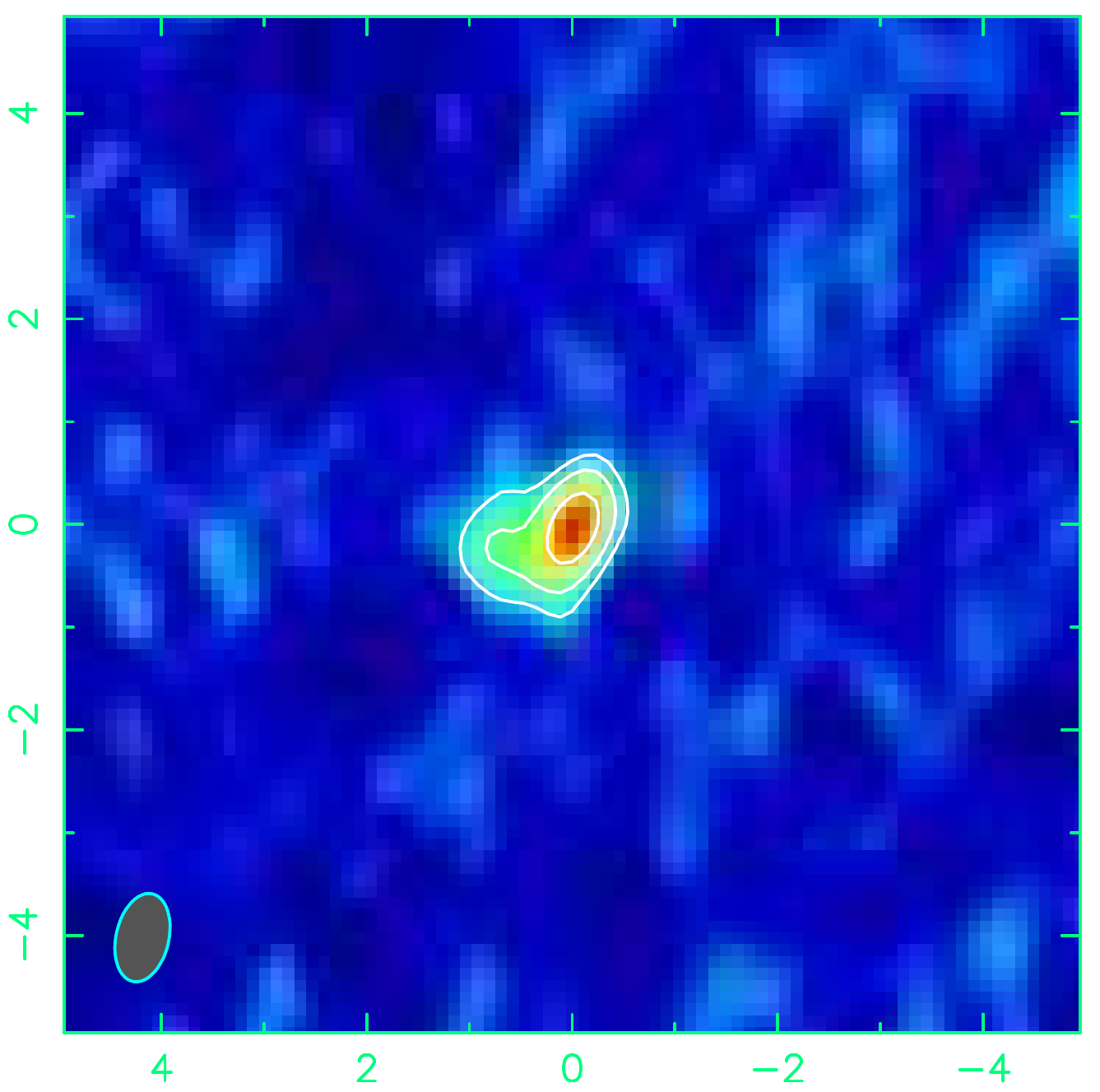}
		\put(12,90){\small\color{white}{HD199178}}
        \put(70,90){\small\color{white}{2020-11-05}}
        \put(45,10){\small\color{white}{Calibrator: J2102+4702}}
        \end{overpic}
		\label{HD_row2}

	\centering
		\centering
		\begin{overpic}[width=0.30\linewidth]{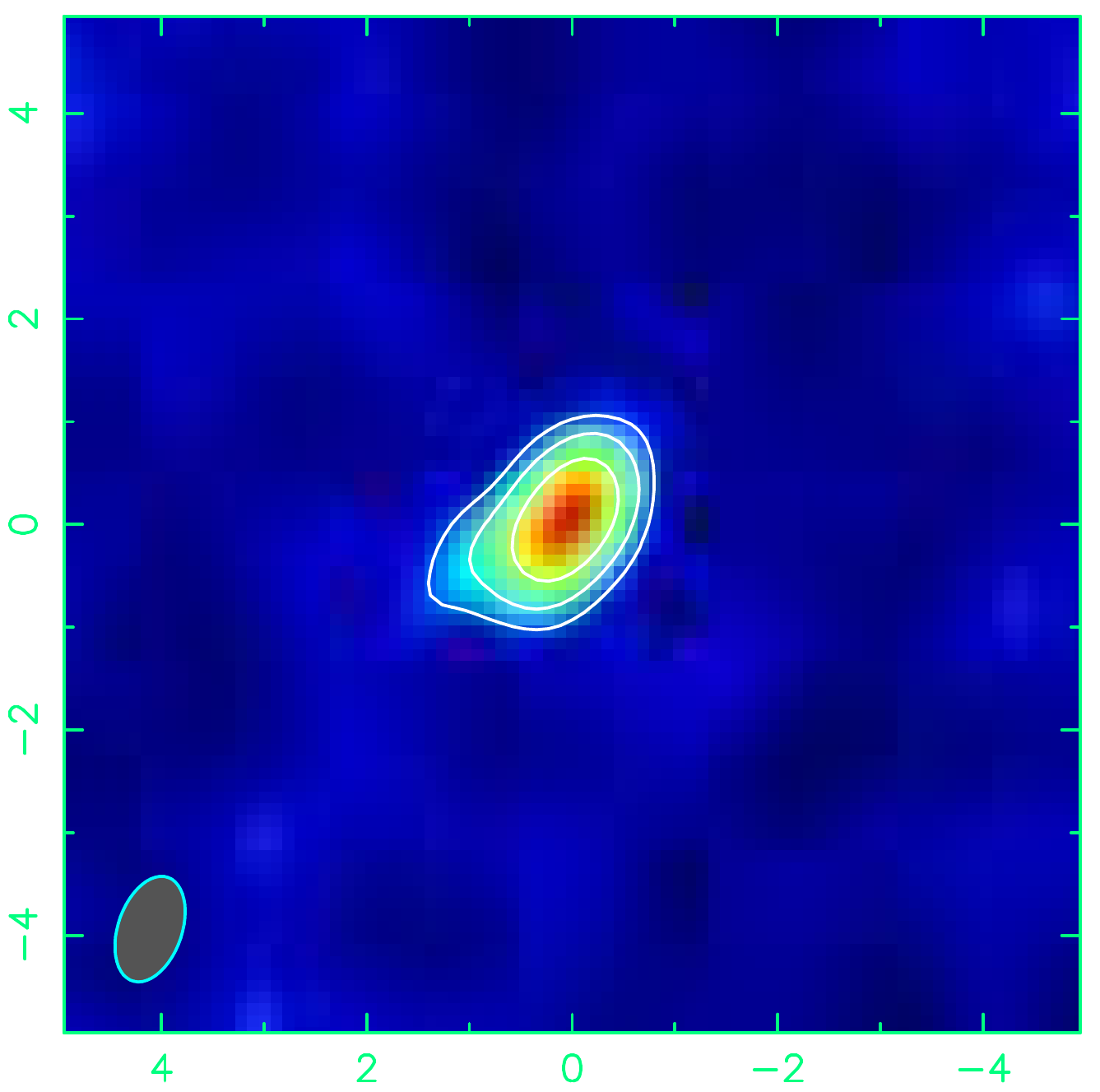}
		\put(12,90){\small\color{white}{HD199178}}
        \put(70,90){\small\color{white}{2020-11-09}}
        \put(45,10){\small\color{white}{Calibrator: J2040+4527}}
        \end{overpic}
        \begin{overpic}[width=0.30\linewidth]{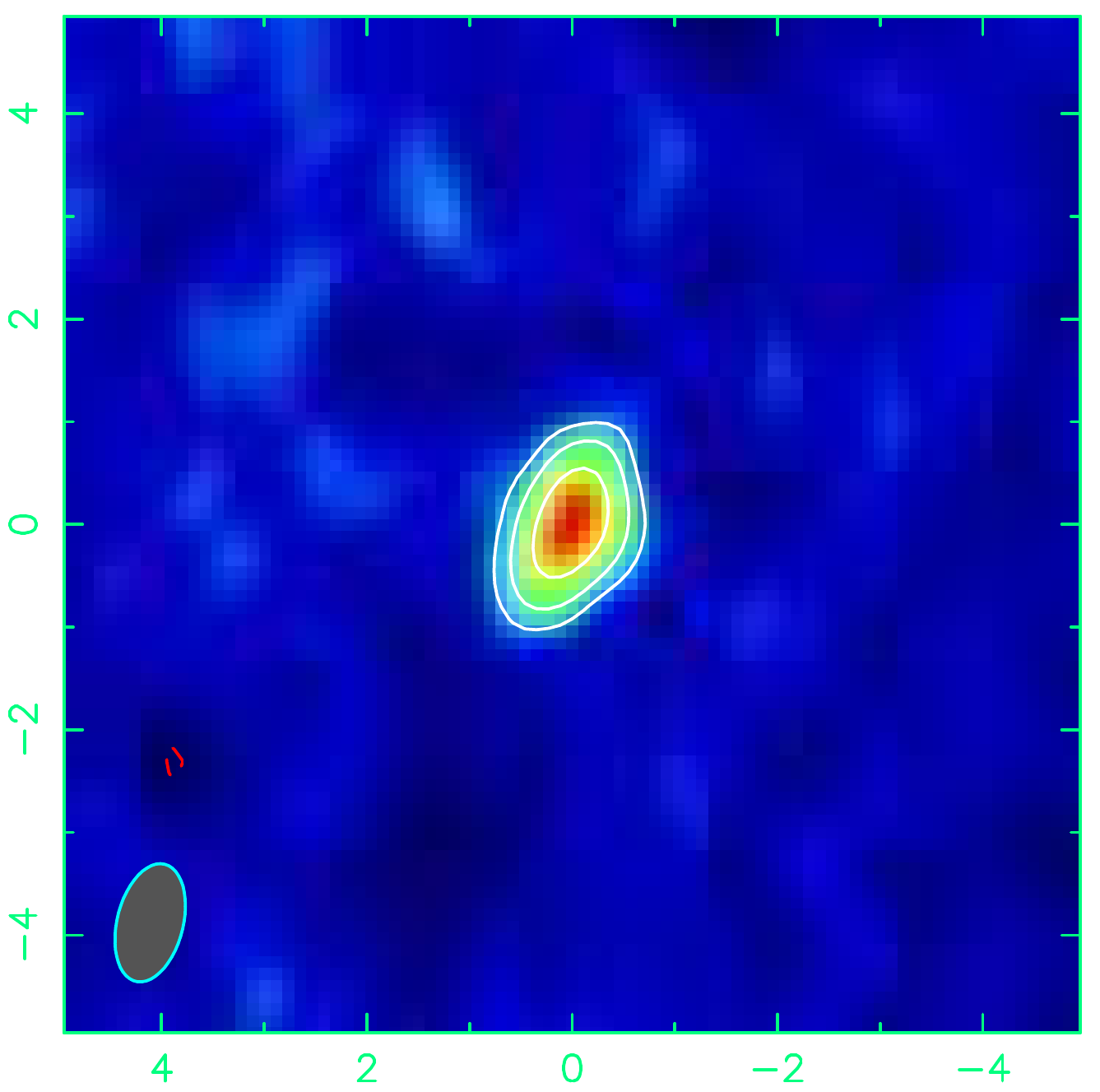}
		\put(12,90){\small\color{white}{HD199178}}
        \put(70,90){\small\color{white}{2021-05-09}}
        \put(45,10){\small\color{white}{Calibrator: J2040+4527}}
        \end{overpic}
        \begin{overpic}[width=0.30\linewidth]{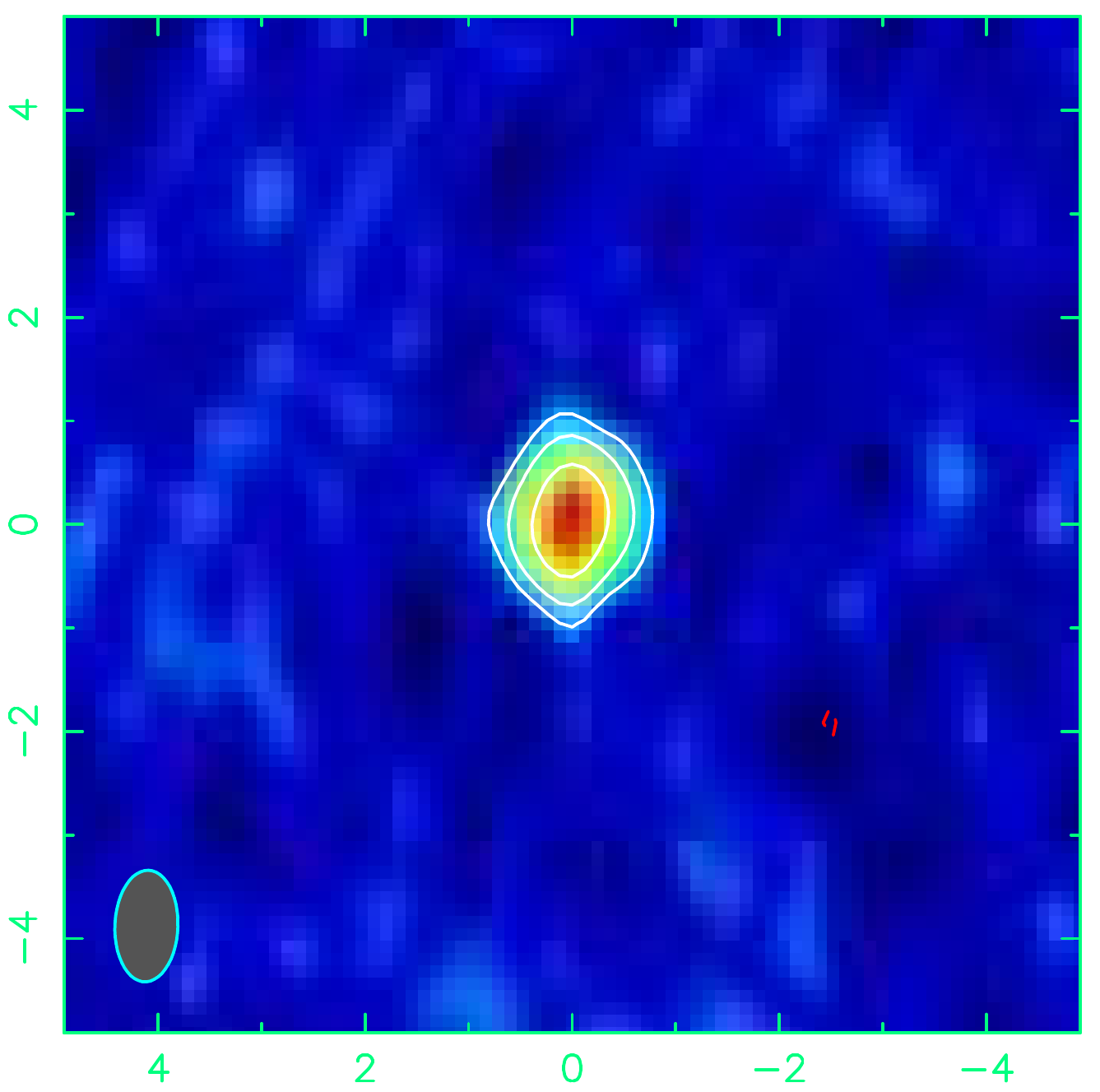}
		\put(12,90){\small\color{white}{HD199178}}
        \put(70,90){\small\color{white}{2021-06-08}}
        \put(45,10){\small\color{white}{Calibrator: J2040+4527}}
        \end{overpic}		
		\label{HD_row3}
	
	\centering
		\centering
		\begin{overpic}[width=0.30\linewidth]{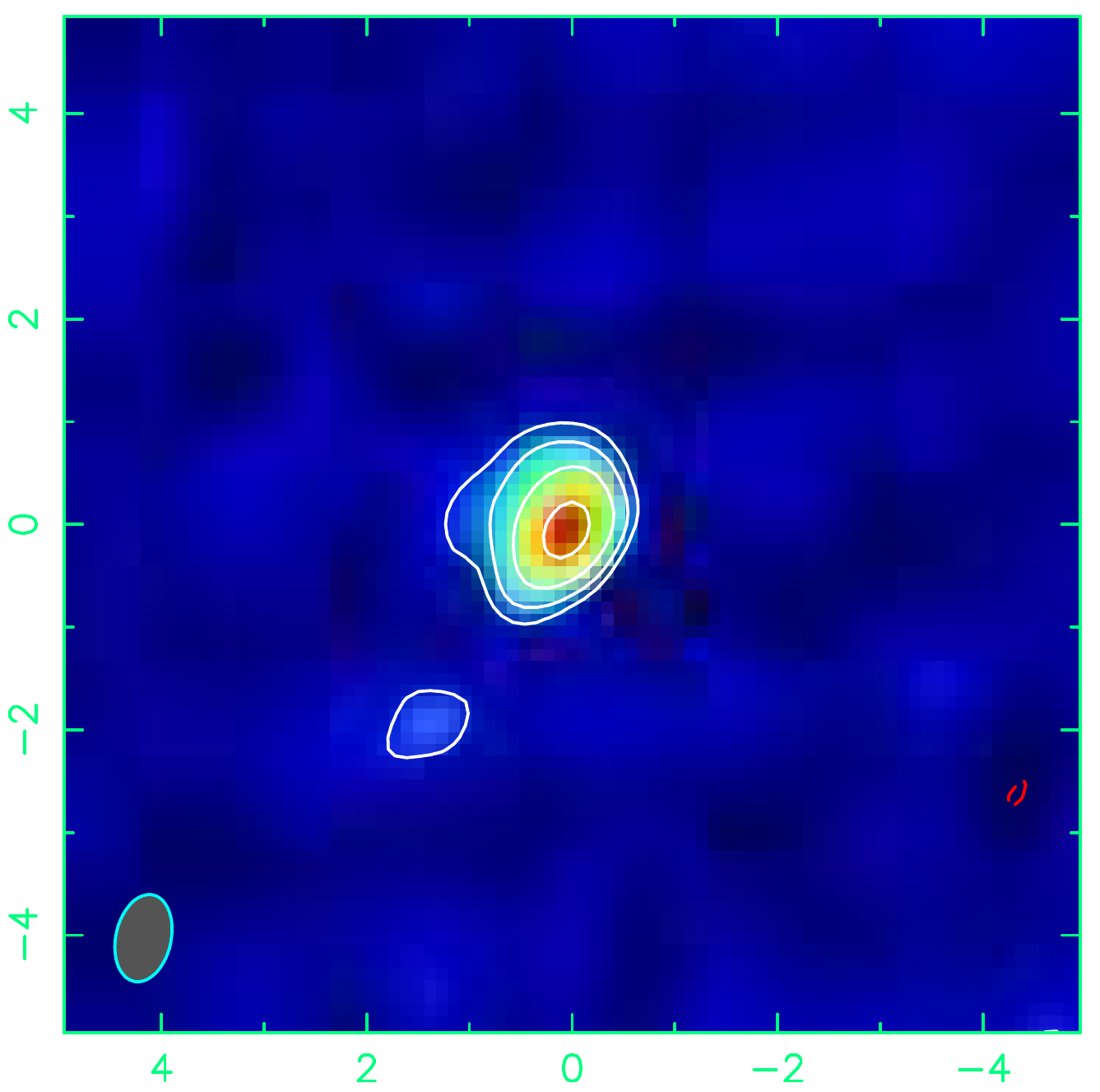}
		\put(12,90){\small\color{white}{HD199178}}
        \put(70,90){\small\color{white}{2020-11-09}}
        \put(45,10){\small\color{white}{Calibrator: J2102+4702}}
        \end{overpic}
        \begin{overpic}[width=0.30\linewidth]{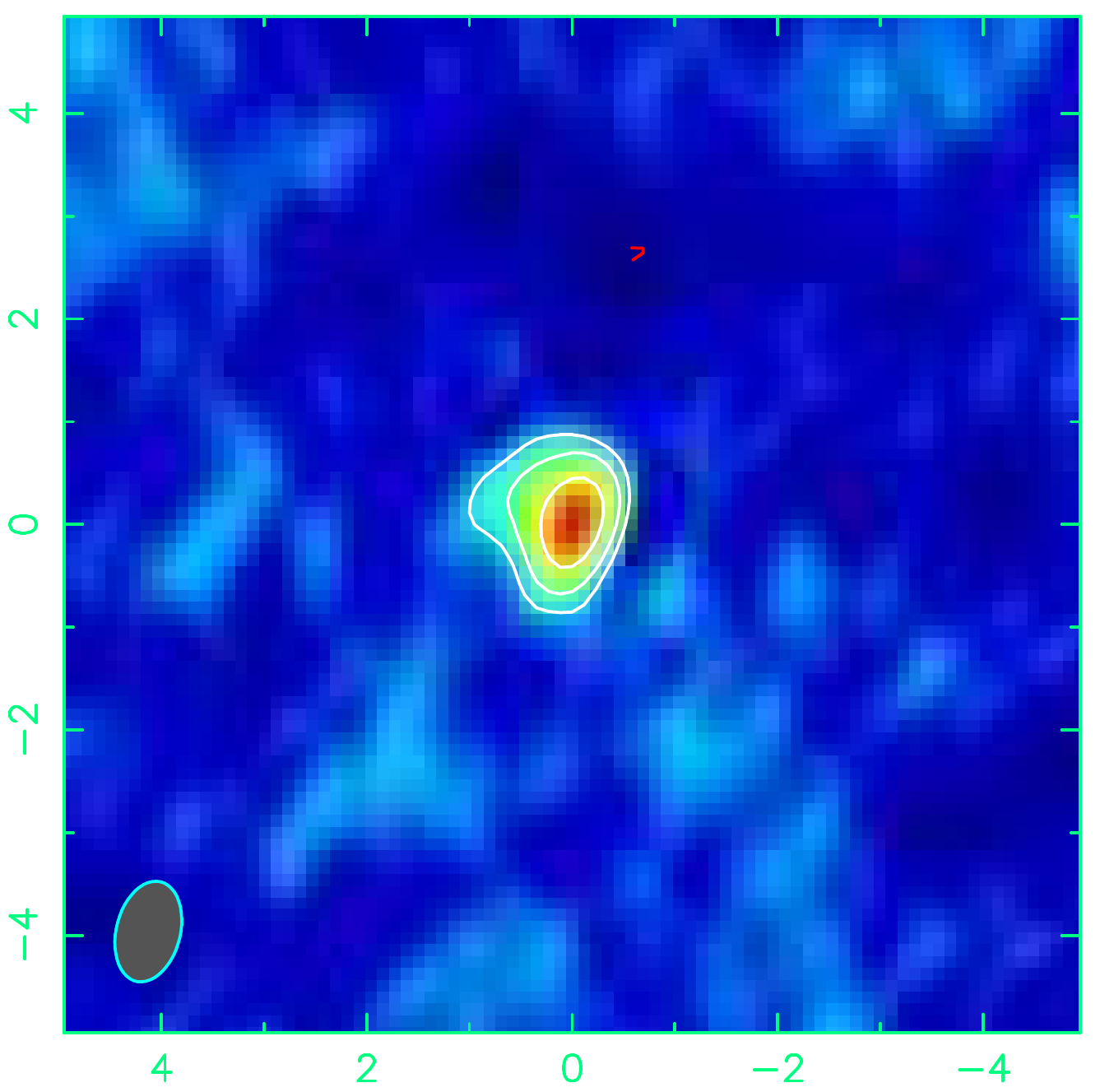}
		\put(12,90){\small\color{white}{HD199178}}
        \put(70,90){\small\color{white}{2021-05-09}}
        \put(45,10){\small\color{white}{Calibrator: J2102+4702}}
        \end{overpic}
        \begin{overpic}[width=0.30\linewidth]{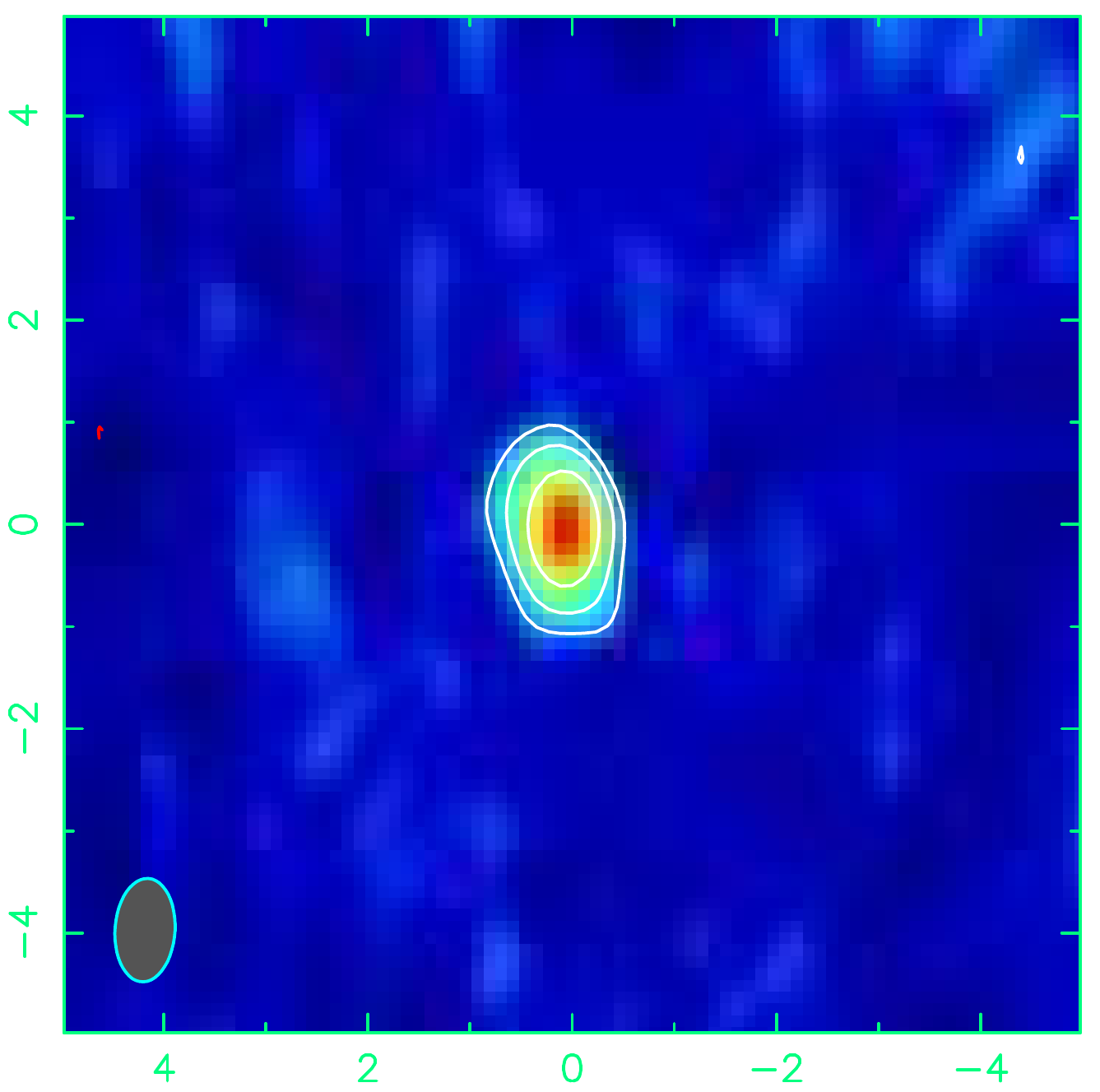}
		\put(12,90){\small\color{white}{HD199178}}
        \put(70,90){\small\color{white}{2021-06-08}}
        \put(45,10){\small\color{white}{Calibrator: J2102+4702}}
        \end{overpic}
		\label{HD_row4}	
	
	\caption{The first and third rows are the \textsc{clean} maps of HD~199178 corresponding to calibrator J2040+4527, while the second and fourth rows are the images of HD~199178 corresponding to calibrator J2102+4702. Image axes are offset from image centre in RA and Dec in mas. The synthesised beam is shown in the bottom left corner of each image. Contours start at three times of the noise level of images and increased by factors of 2.}
	\label{HD_imgs}
\end{figure*}

\begin{figure*}
    \centering
    \centering
        \begin{overpic}[width=0.30\linewidth]{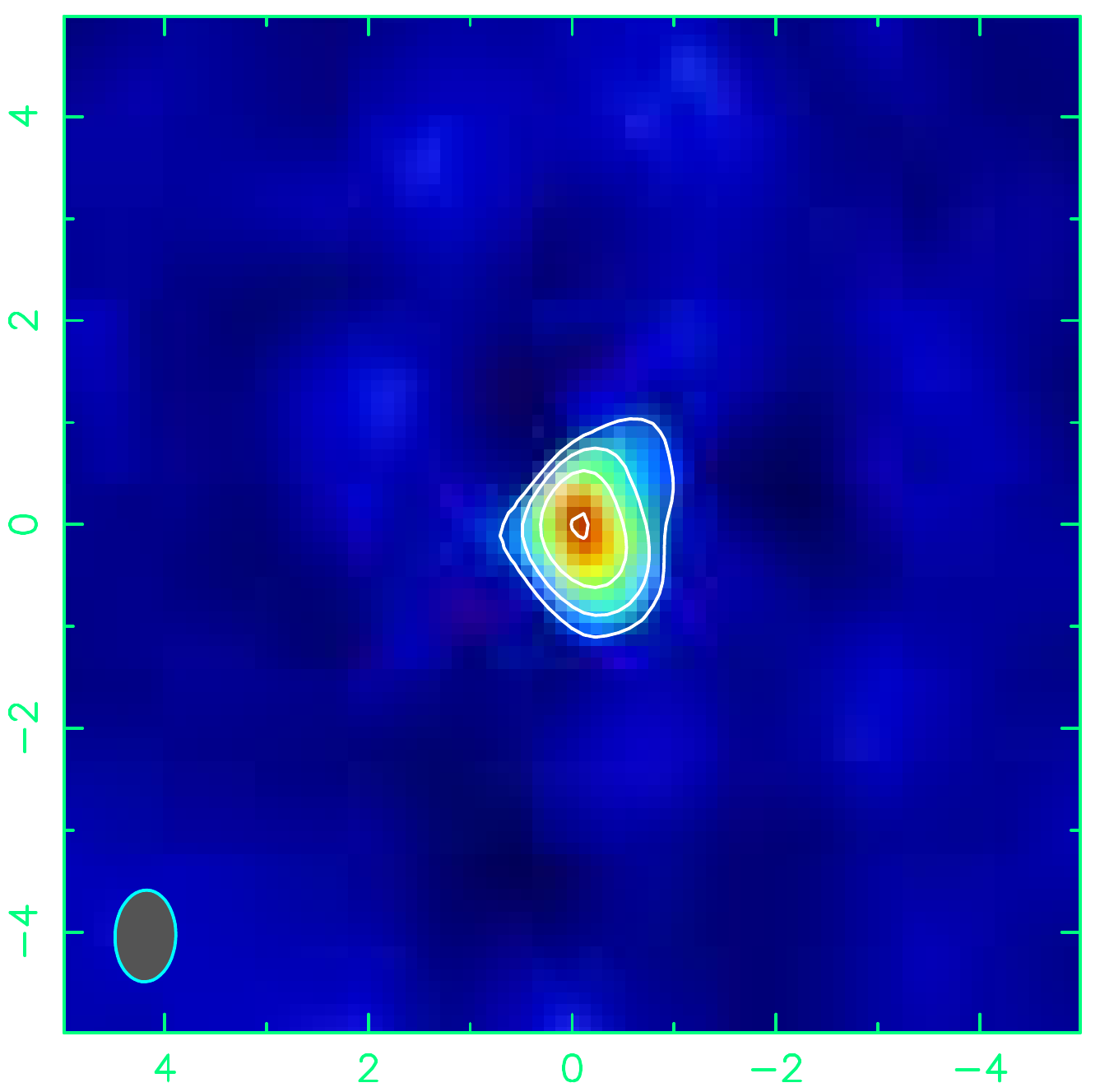}
		\put(12,90){\small\color{white}{AR Lac}}
            \put(70,90){\small\color{white}{2020-05-09}}
            \put(45,10){\small\color{white}{Calibrator: J2153+4322}}
        \end{overpic}
        \begin{overpic}[width=0.30\linewidth]{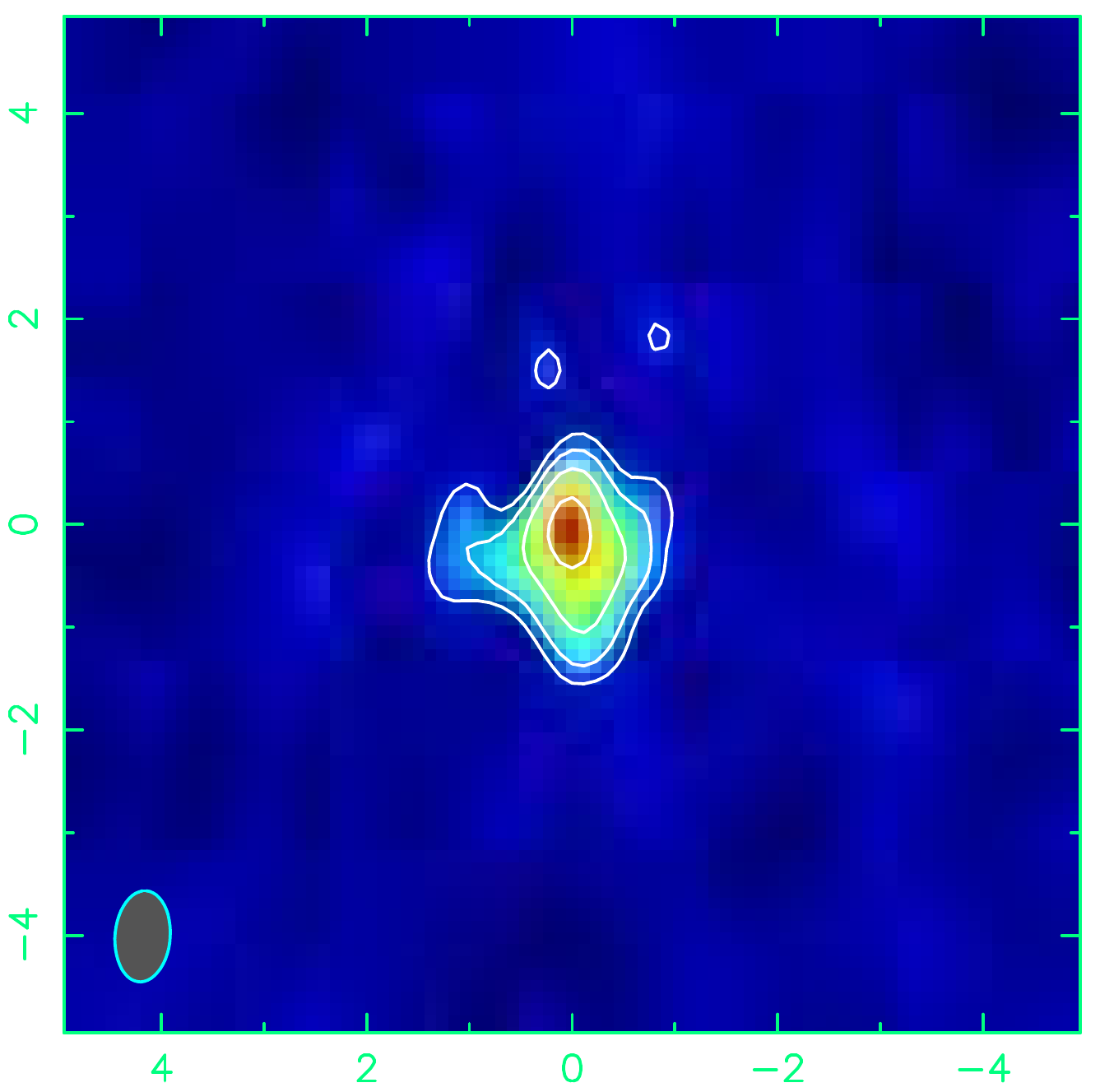}
		\put(12,90){\small\color{white}{AR Lac}}
            \put(70,90){\small\color{white}{2020-05-17}}
            \put(45,10){\small\color{white}{Calibrator: J2153+4322}}
        \end{overpic}
        \begin{overpic}[width=0.30\linewidth]{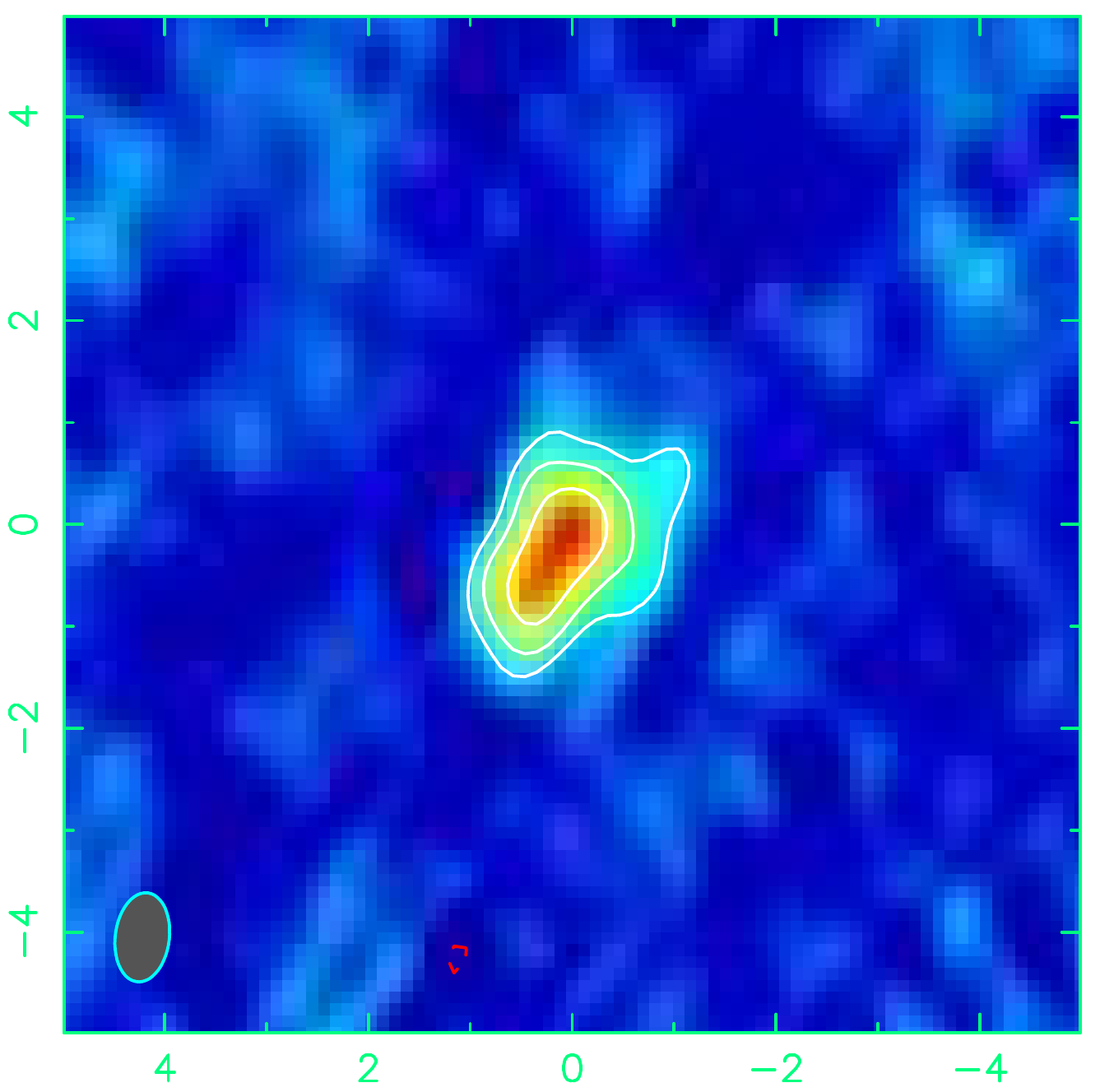}
		\put(12,90){\small\color{white}{AR Lac}}
            \put(70,90){\small\color{white}{2020-11-05}}
            \put(45,10){\small\color{white}{Calibrator: J2153+4322}}
        \end{overpic}
    \label{AR_row1}

    \centering
    \centering
        \begin{overpic}[width=0.30\linewidth]{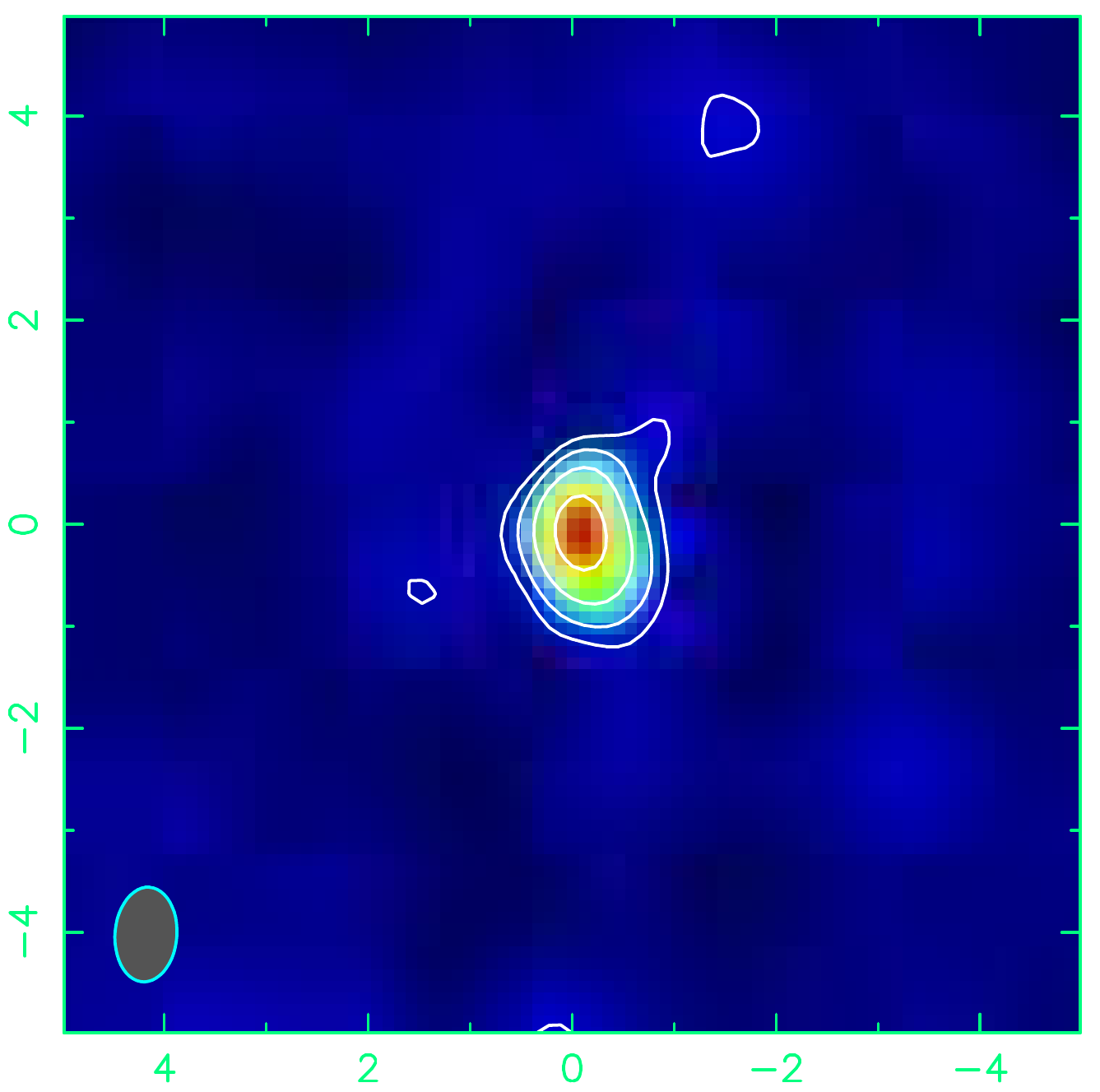}
	    \put(12,90){\small\color{white}{AR Lac}}
            \put(70,90){\small\color{white}{2020-05-09}}
            \put(45,10){\small\color{white}{Calibrator: J2202+4216}}
        \end{overpic}
        \begin{overpic}[width=0.30\linewidth]{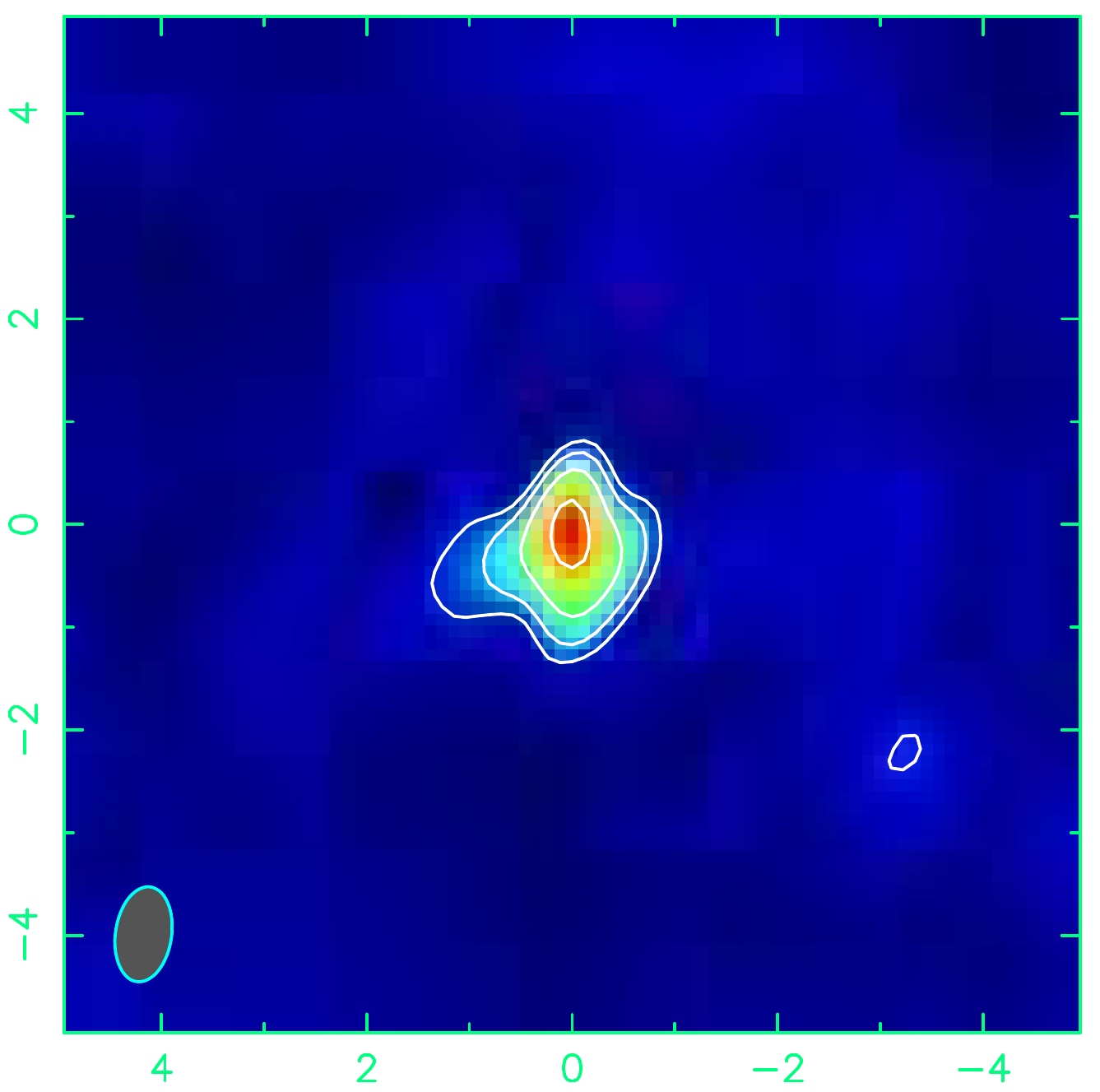}
	    \put(12,90){\small\color{white}{AR Lac}}
            \put(70,90){\small\color{white}{2020-05-17}}
            \put(45,10){\small\color{white}{Calibrator: J2202+4216}}
        \end{overpic}
        \begin{overpic}[width=0.30\linewidth]{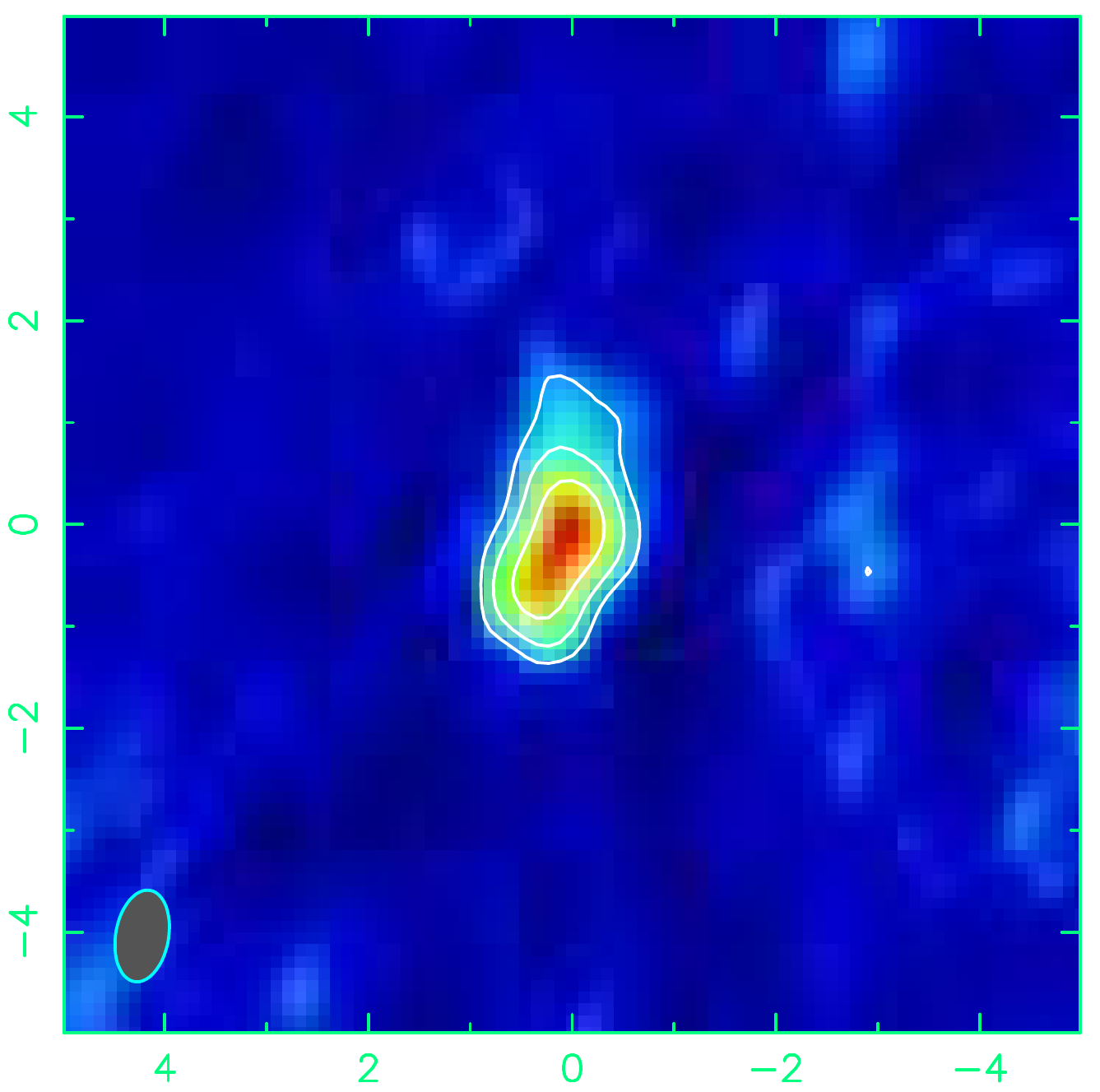}
	    \put(12,90){\small\color{white}{AR Lac}}
            \put(70,90){\small\color{white}{2020-11-05}}
            \put(45,10){\small\color{white}{Calibrator: J2202+4216}}
        \end{overpic}		
    \label{AR_row2}
	
    \centering
	\centering
	\begin{overpic}[width=0.30\linewidth]{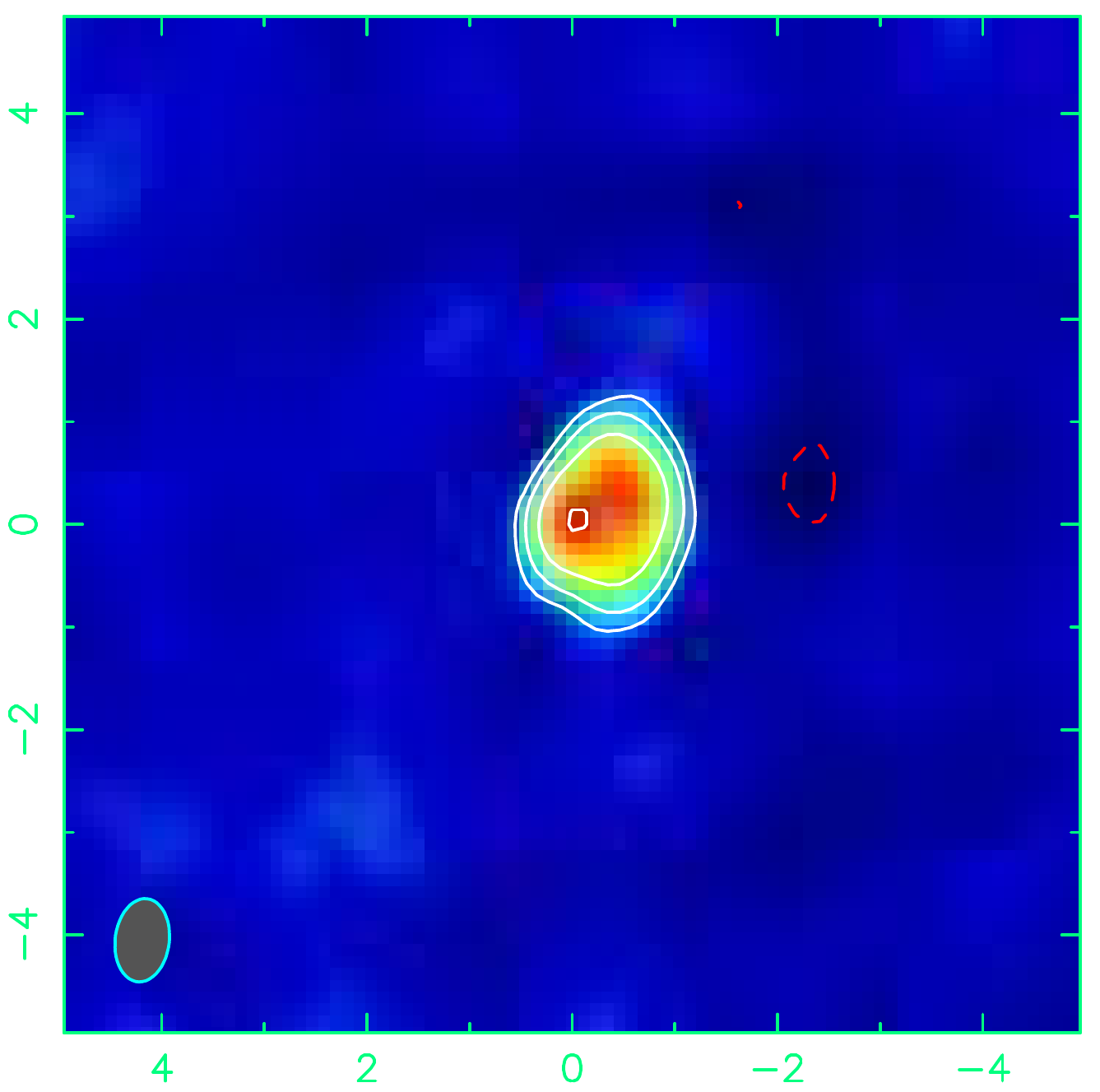}
		\put(12,90){\small\color{white}{AR Lac}}
            \put(70,90){\small\color{white}{2020-11-09}}
            \put(45,10){\small\color{white}{Calibrator: J2153+4322}}
        \end{overpic}
        \begin{overpic}[width=0.30\linewidth]{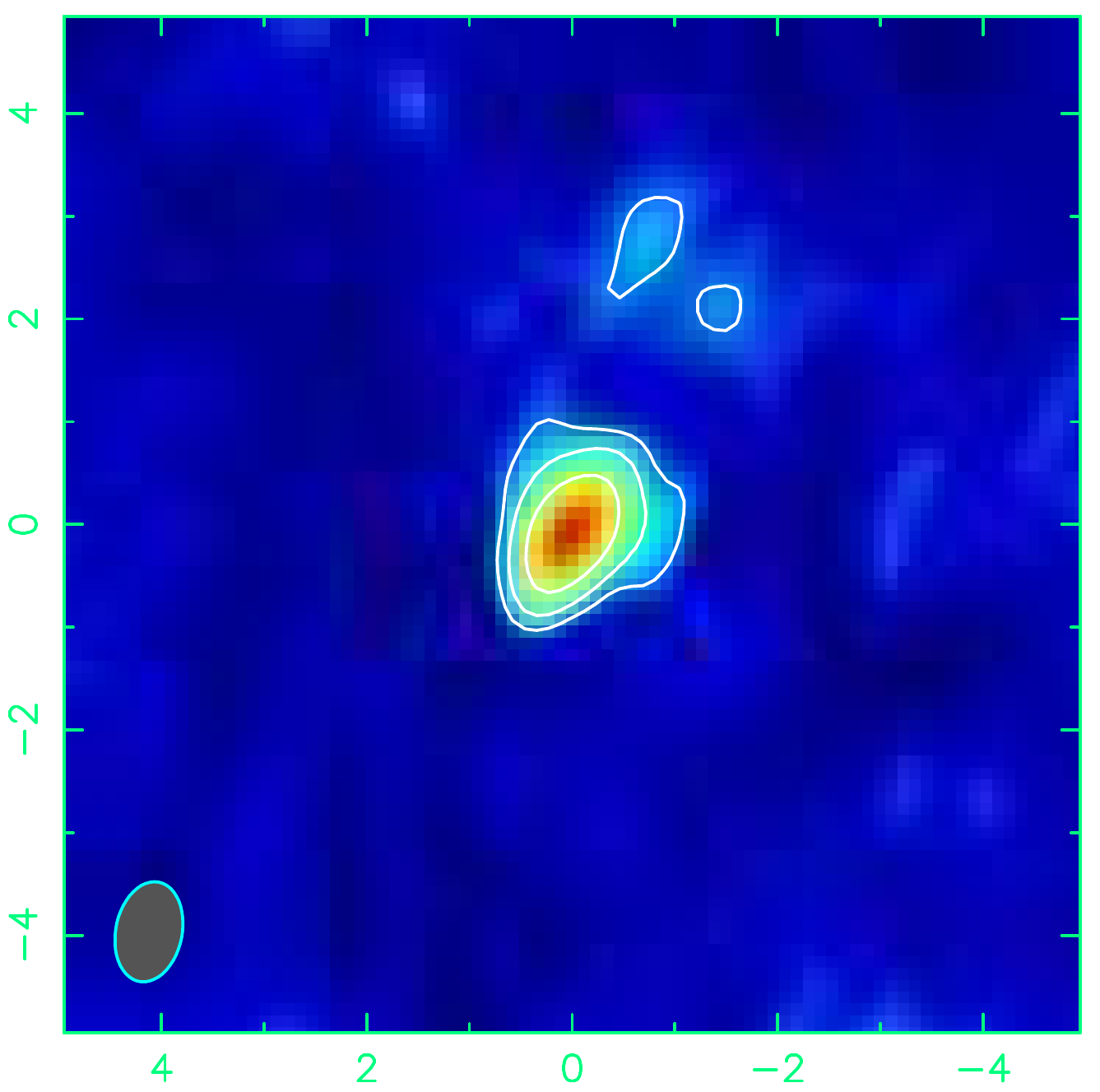}
		\put(12,90){\small\color{white}{AR Lac}}
            \put(70,90){\small\color{white}{2021-05-09}}
            \put(45,10){\small\color{white}{Calibrator: J2153+4322}}
        \end{overpic}
        \begin{overpic}[width=0.30\linewidth]{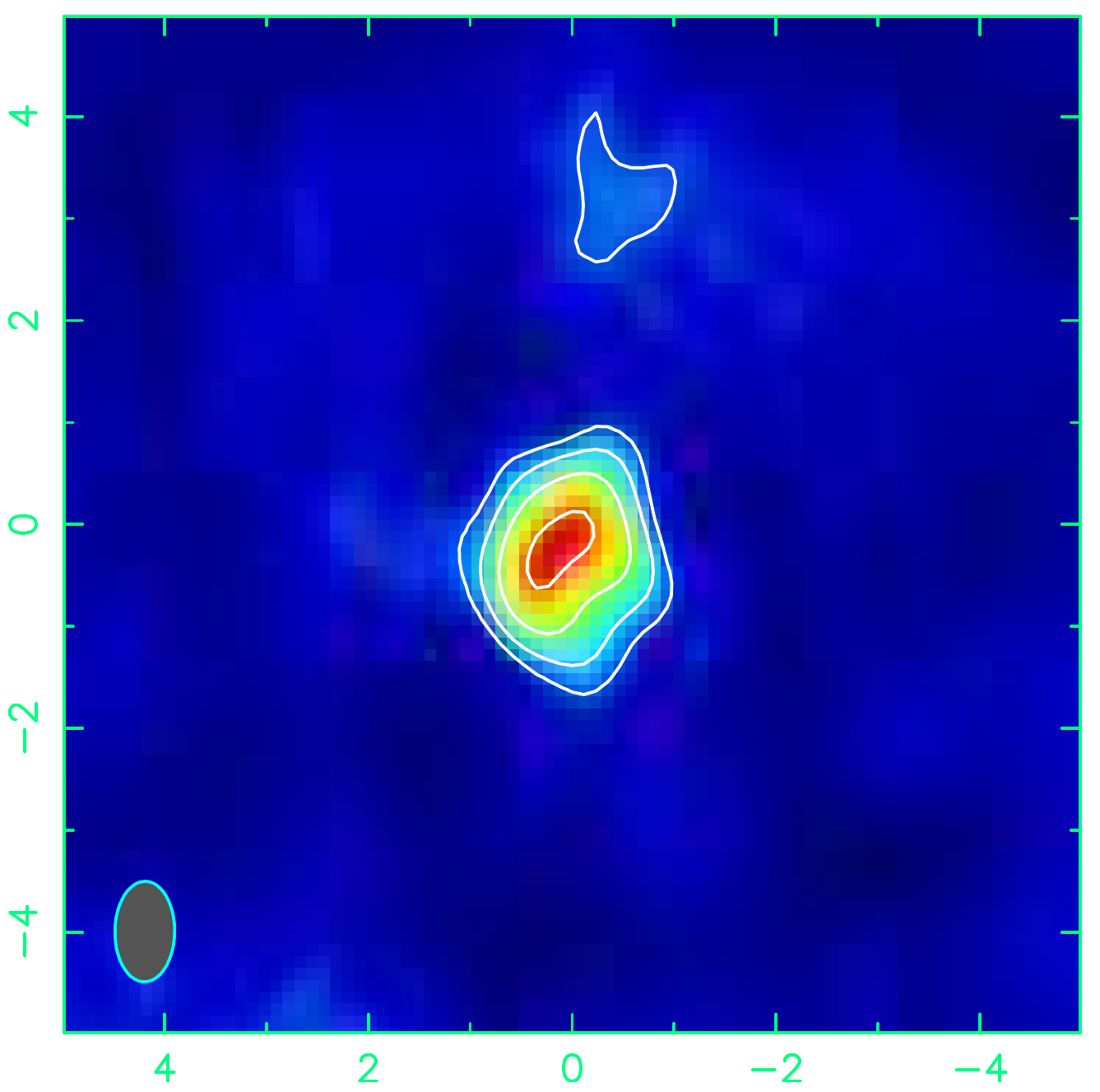}
		\put(12,90){\small\color{white}{AR Lac}}
            \put(70,90){\small\color{white}{2021-06-08}}
            \put(45,10){\small\color{white}{Calibrator: J2153+4322}}
        \end{overpic}		
	\label{AR_row3}
	
    \centering
	\centering
	\begin{overpic}[width=0.30\linewidth]{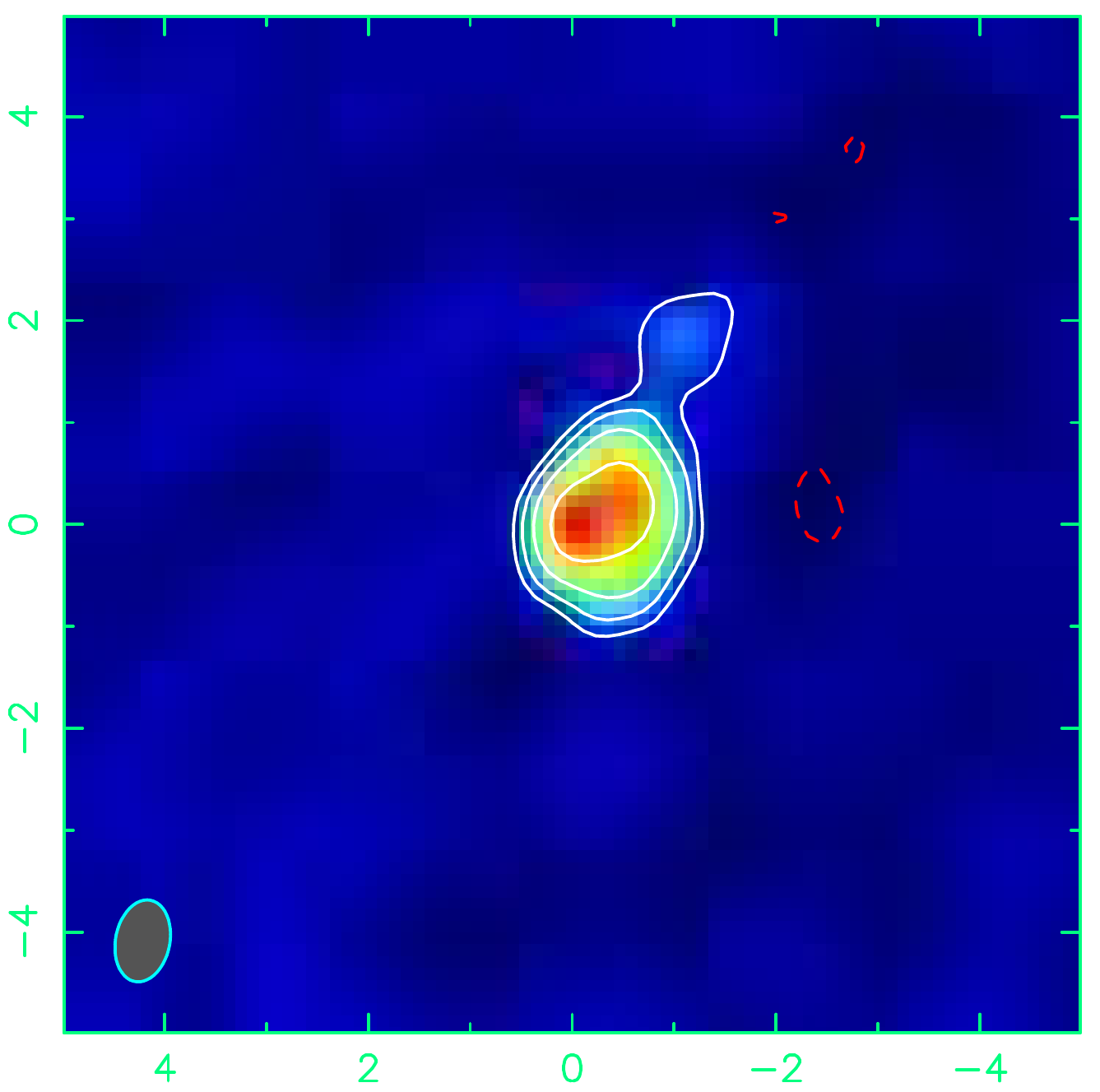}
		\put(12,90){\small\color{white}{AR Lac}}
            \put(70,90){\small\color{white}{2020-11-09}}
            \put(45,10){\small\color{white}{Calibrator: J2202+4216}}
        \end{overpic}
        \begin{overpic}[width=0.30\linewidth]{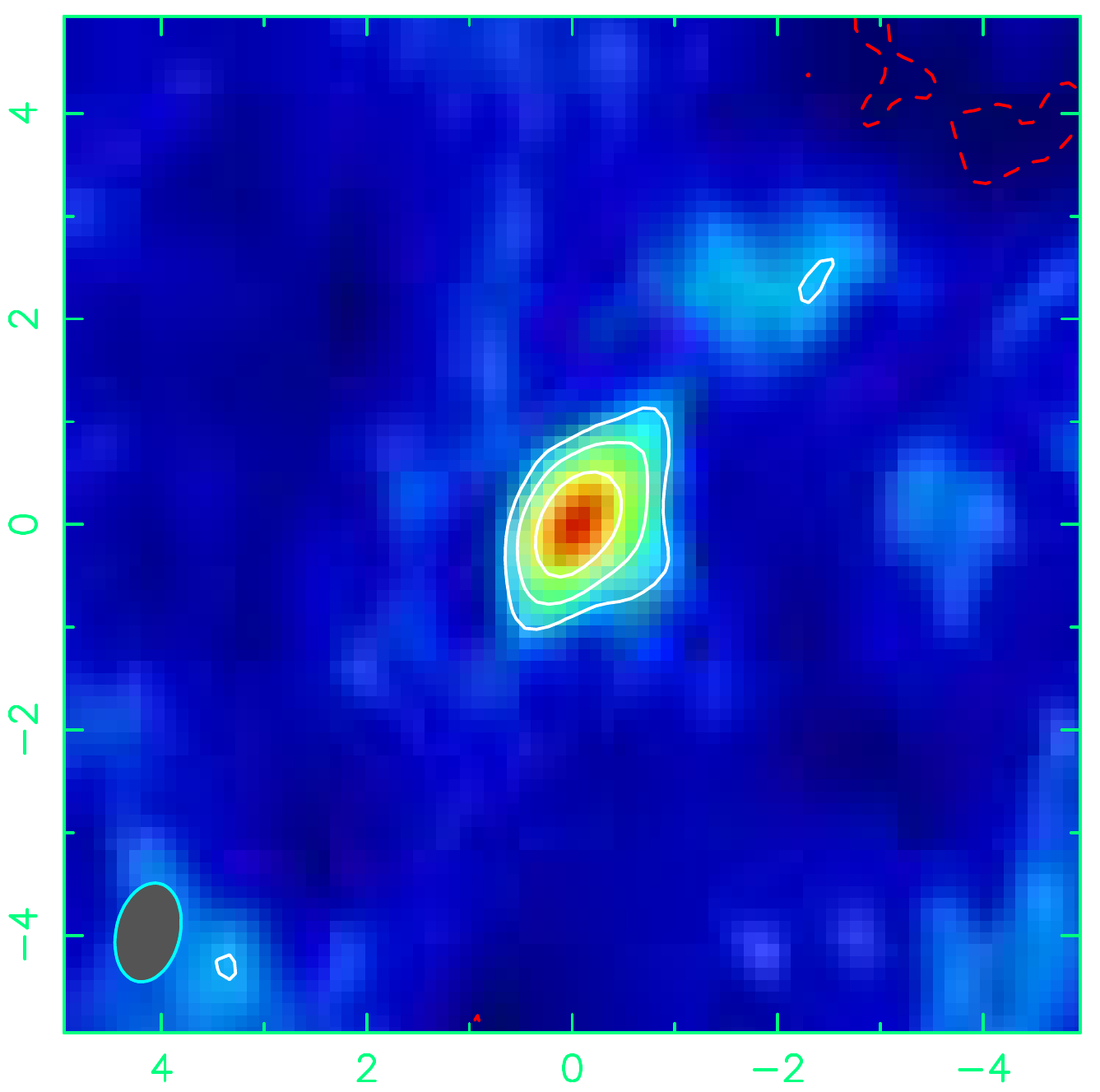}
		\put(12,90){\small\color{white}{AR Lac}}
            \put(70,90){\small\color{white}{2021-05-09}}
            \put(45,10){\small\color{white}{Calibrator: J2202+4216}}
        \end{overpic}
        \begin{overpic}[width=0.30\linewidth]{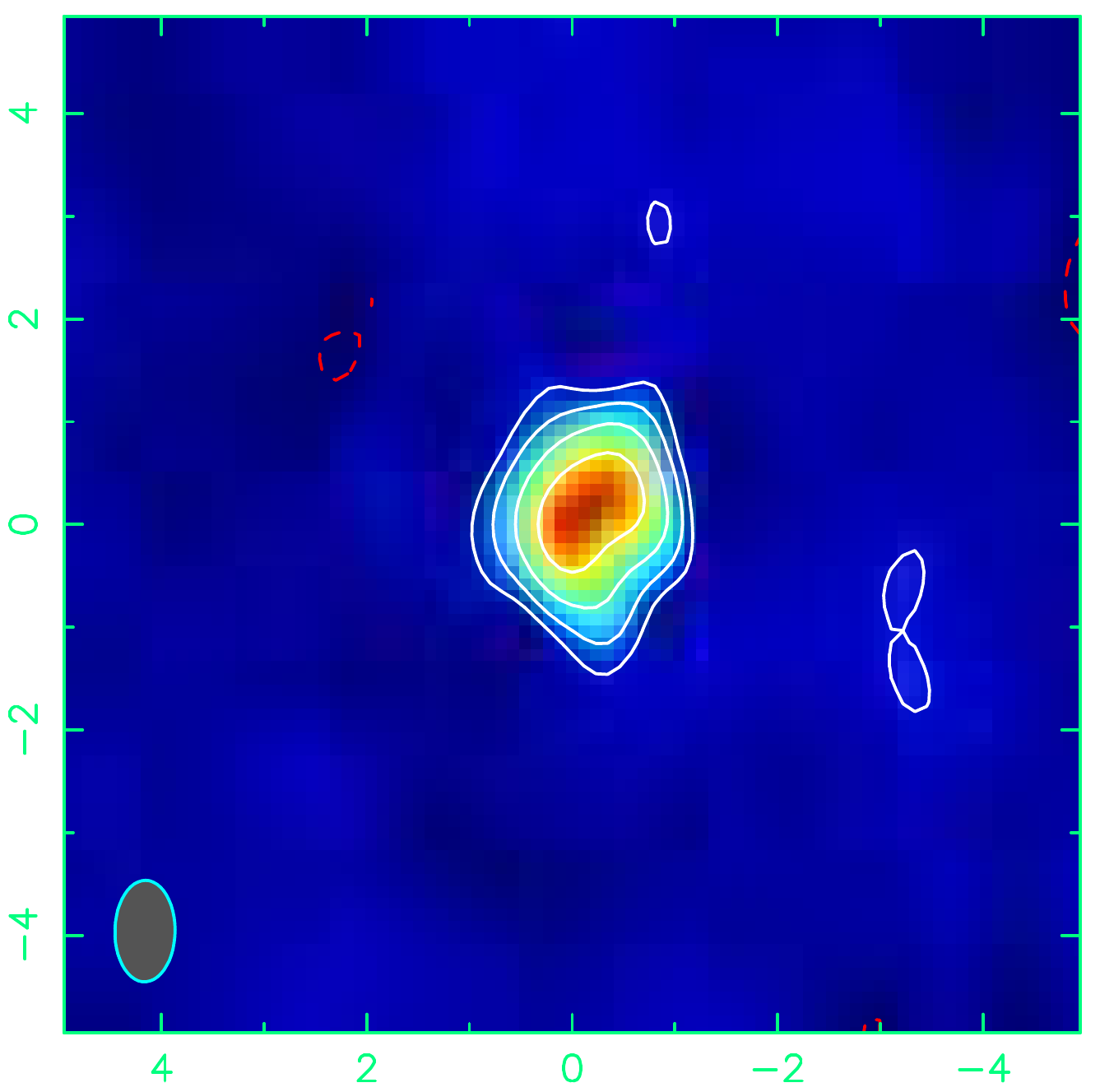}
		\put(12,90){\small\color{white}{AR Lac}}
            \put(70,90){\small\color{white}{2021-06-08}}
            \put(45,10){\small\color{white}{Calibrator: J2202+4216}}
        \end{overpic}
	\label{AR_row4}	
\caption{The first and third rows are the \textsc{clean} maps of AR~Lac corresponding to calibrator J2153+4322, while the second and fourth rows are the images of AR Lac corresponding to calibrator J2202+4216. Image axes are offset from image centre in RA and Dec in mas. The synthesised beam is shown in the bottom left corner of each image. Contours start at three times of the noise level of images and increased by factors of 2.}
\label{ARLac_imgs}
\end{figure*}

\begin{figure*}
	\centering
		\centering
        \begin{overpic}[width=0.24\linewidth]{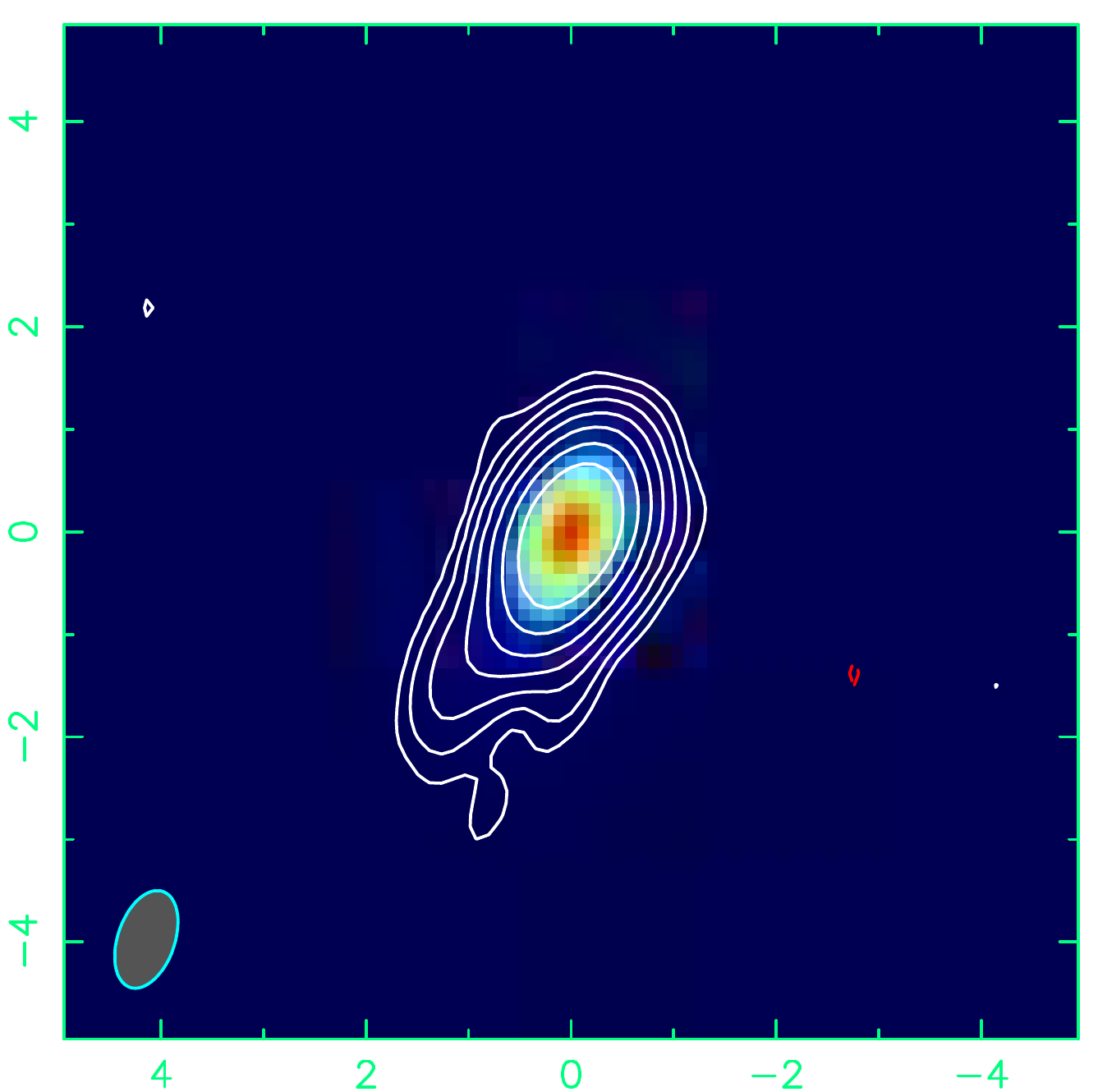} 
        \put(12,90){\scriptsize\color{white}{J2040+4527}}
        \put(55,13){\scriptsize\color{white}{rms: 0.077 mJy/beam}}
        \put(55,9){\scriptsize\color{white}{Peak: 0.0471 Jy/beam}}
        \put(16.5,13.5){\scriptsize\color{white}{0.99 $\times$ 0.557 mas}}
        \put(16,9){\scriptsize\color{white}{@ --19$^\circ$}}
        \end{overpic}
        \begin{overpic}[width=0.24\linewidth]{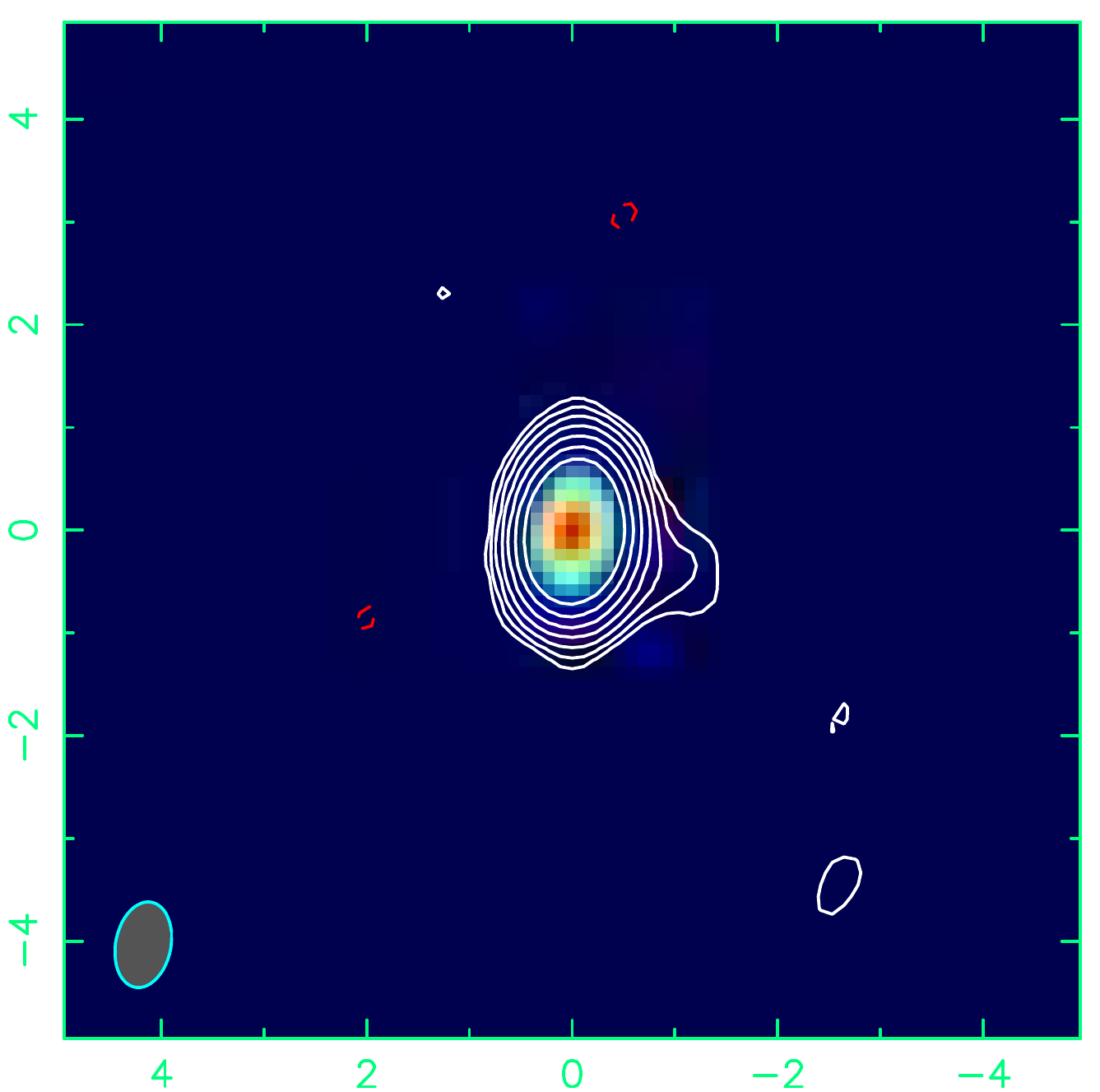}
        \put(12,90){\scriptsize\color{white}{J2102+4702}}
        \put(55,12){\scriptsize\color{white}{rms: 0.061 mJy/beam}}
        \put(55,8){\scriptsize\color{white}{Peak: 0.0883 Jy/beam}}
        \put(16.5,13.5){\scriptsize\color{white}{0.843 $\times$ 0.538 mas}}
        \put(16,9){\scriptsize\color{white}{@ --11.1$^\circ$}}
        \end{overpic}
        \begin{overpic}[width=0.24\linewidth]{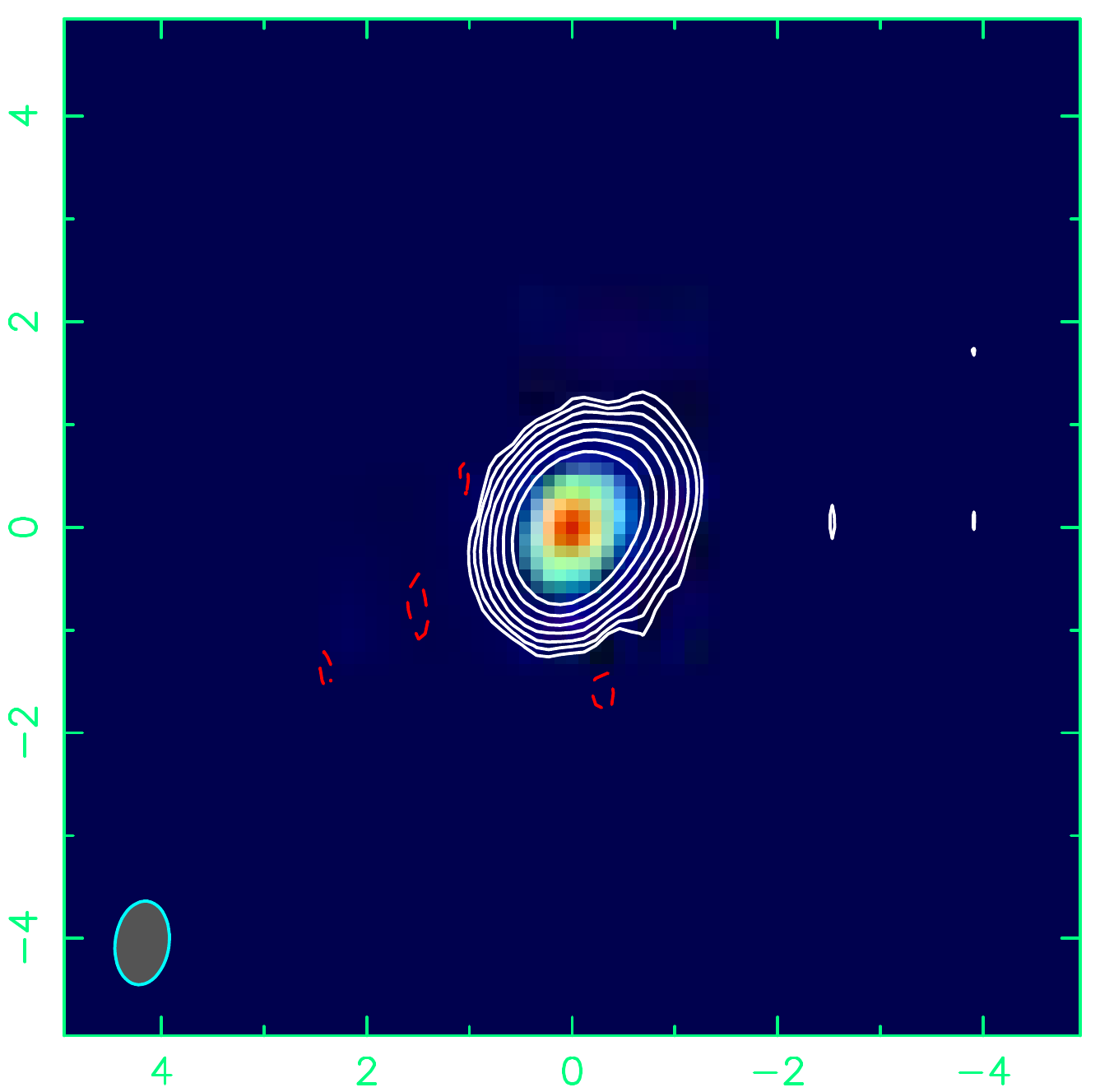}
        \put(12,90){\scriptsize\color{white}{J2153+4322}}
        \put(55,13){\scriptsize\color{white}{rms: 0.067 mJy/beam}}
        \put(55,9){\scriptsize\color{white}{Peak: 0.142 Jy/beam}}
        \put(16,13.5){\scriptsize\color{white}{0.817 $\times$ 0.524 mas}}
        \put(16,9){\scriptsize\color{white}{@ --7.41$^\circ$}}
        \end{overpic}
		\begin{overpic}[width=0.237\linewidth]{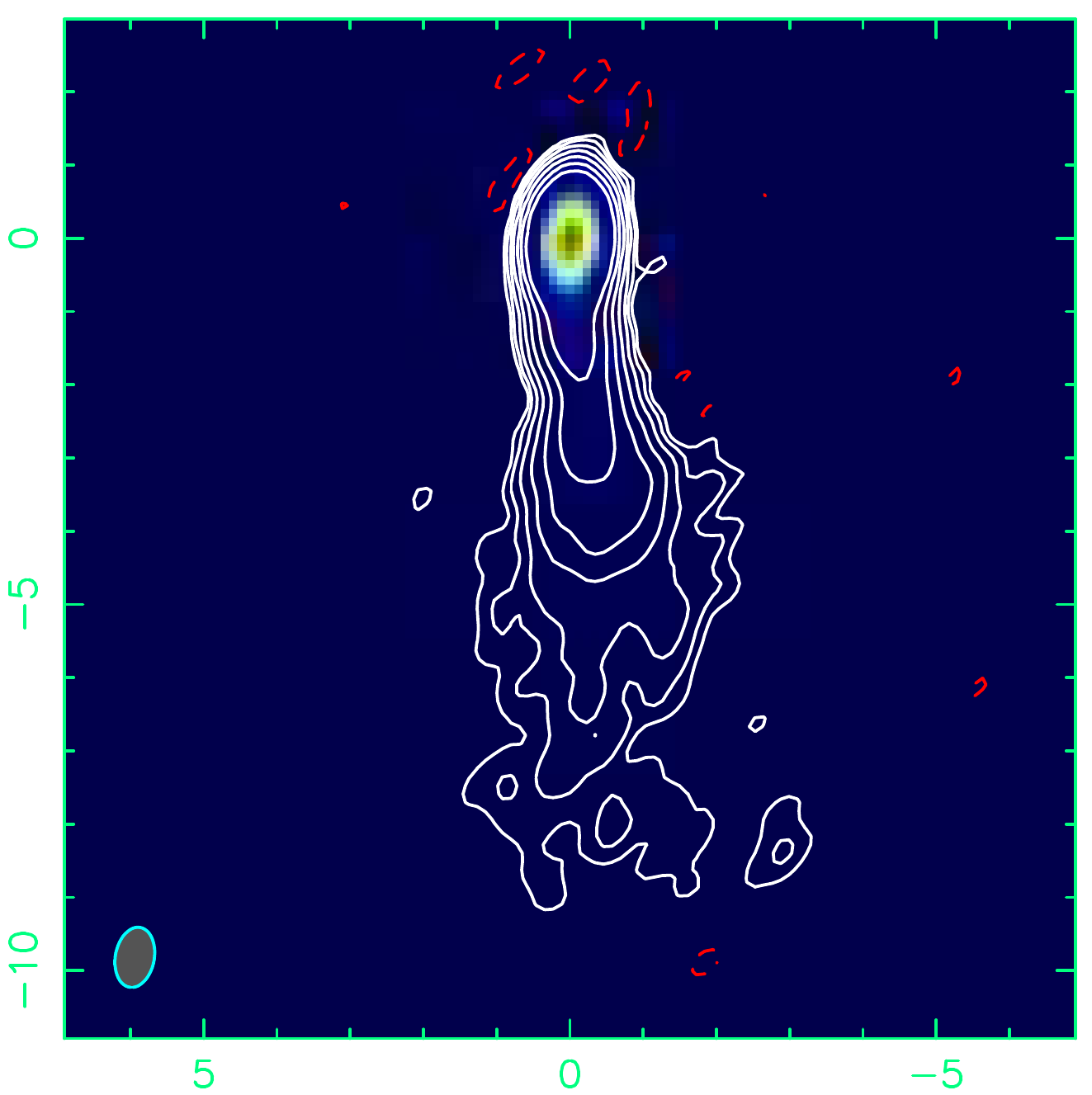}
        \put(12,90){\scriptsize\color{white}{J2202+4216}}
        \put(55,13){\scriptsize\color{white}{rms: 0.246 mJy/beam}}
        \put(55,9){\scriptsize\color{white}{Peak: 1.570 Jy/beam}}
        \put(16,13.5){\scriptsize\color{white}{0.829 $\times$ 0.532 mas}}
        \put(16,9){\scriptsize\color{white}{@ --10.4$^\circ$}}
        \end{overpic}
		\label{row4}
	\caption{Images of the calibrators from VLBA observations on 2020-11-09. The synthesis restoring beam is shown in the bottom left corner of each panel. Contours levels start at three times of the image 3-$\sigma$ noise and increased by factors of 2.}
	\label{calibs_img}
\end{figure*}

The \textsc{clean} maps of  
HD~199178 and AR~Lac at each epoch are displayed in Fig.~\ref{HD_imgs} and \ref{ARLac_imgs}. The relative offsets in RA and Dec are in the unit of mas. The related calibrator is also given in the bottom-right corner. These images have a mean resolution of $0.91 \times 0.57$~mas. Our images reach a sensitivity of 0.053--0.120~mJy\,beam$^{-1}$. Table~\ref{tab:targets_coords} lists the map parameters. With respect to the correlation phase centres, the relative offsets are measured via running the \textsc{aips} task \texttt{JMFIT}. Astrometric parameters of the two radio stars are listed in Table~\ref{tab:targets_fitted_results}. The coordinates of the radio stars relative to their calibrators on 2020 November 9 are listed in Table~\ref{tab:Coords_radiostars} and the images of the calibrators are shown in Fig.~\ref{calibs_img}. The integrated flux densities are reported in Table~\ref{calibs_fluxs} and the light curves are shown in Fig.~\ref{fig:flux_all}.

\begin{figure}
    \includegraphics[width=\columnwidth]{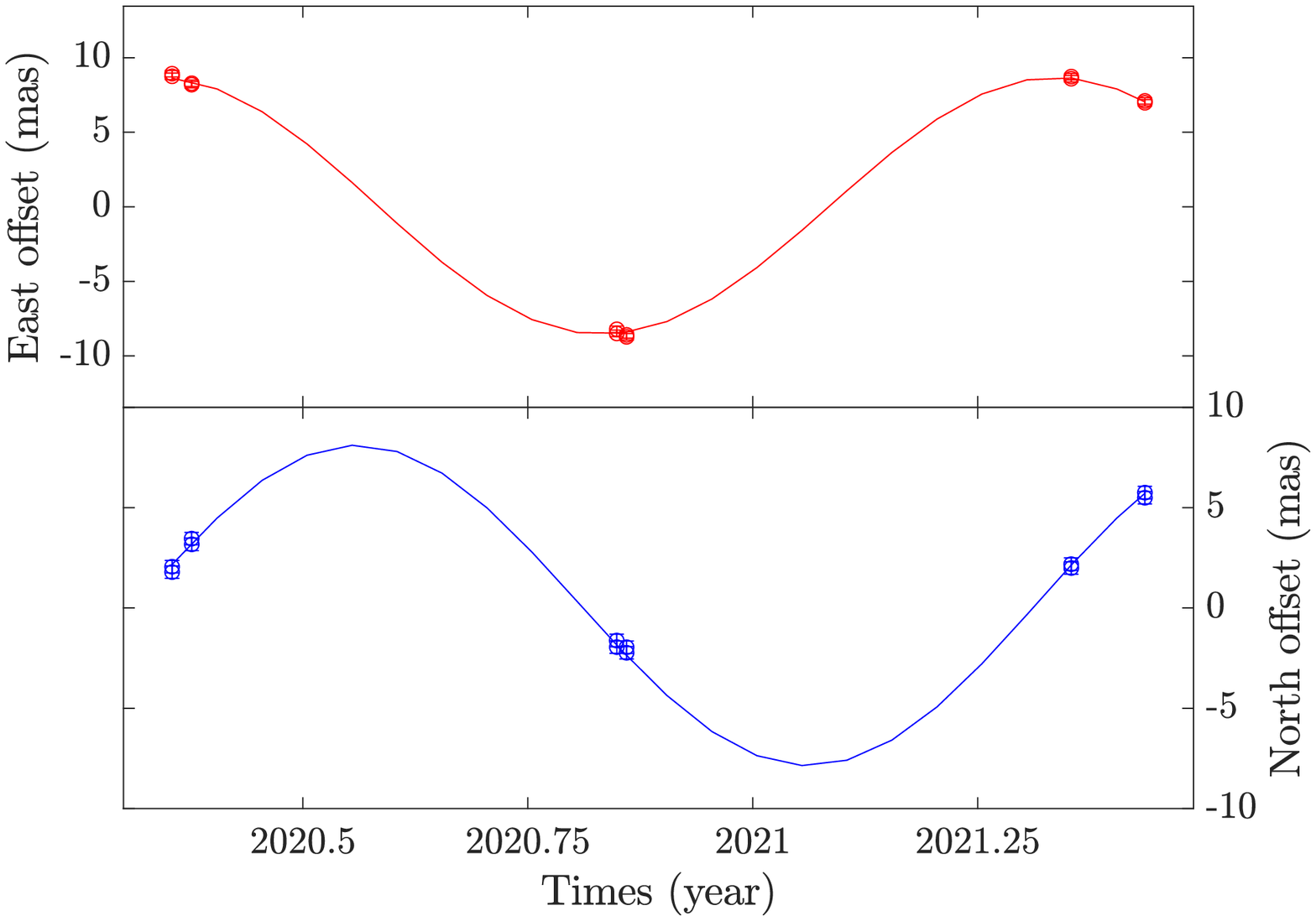}
    \includegraphics[width=\columnwidth]{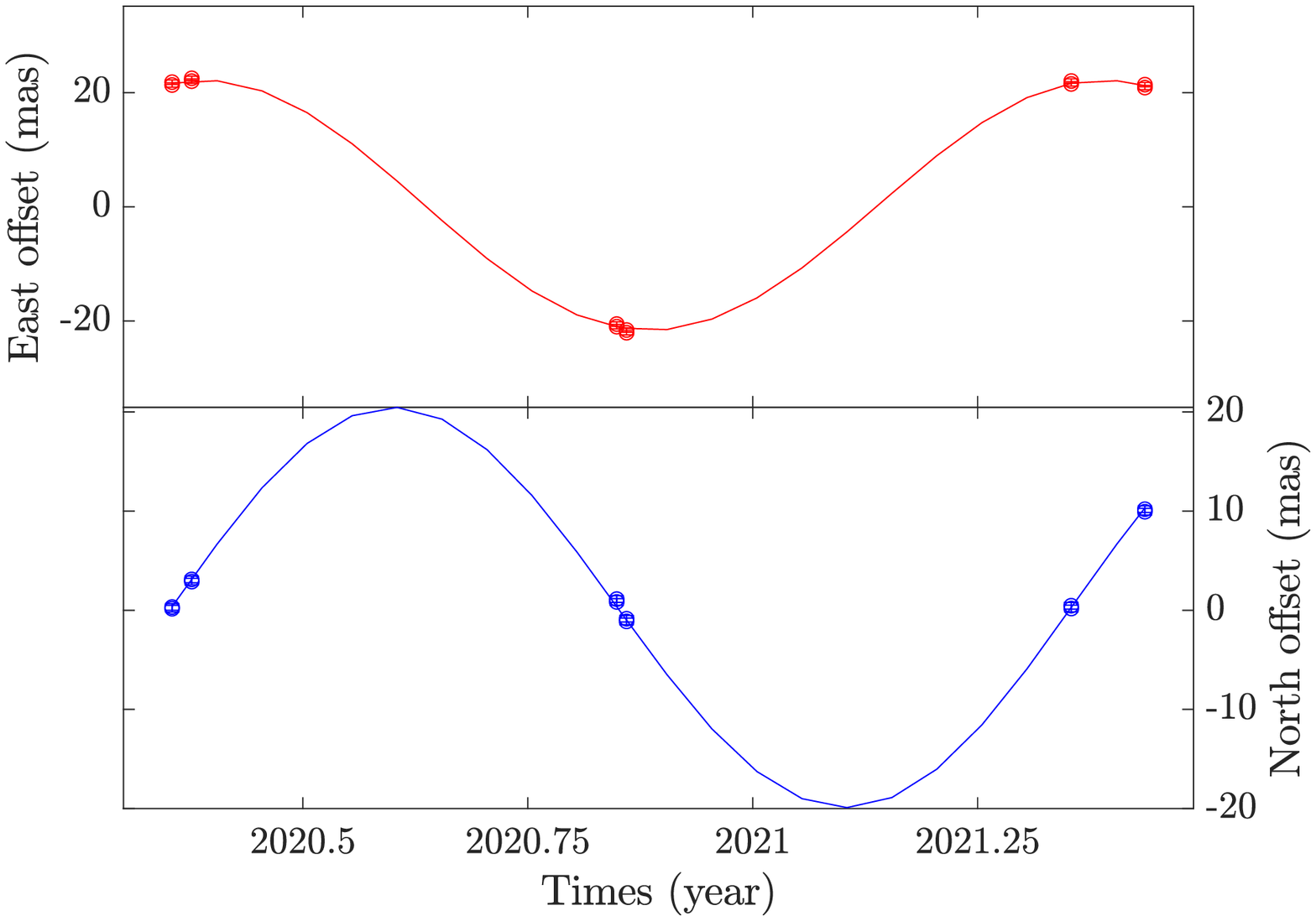}
 \caption{The least-squares astrometric fit for the two radio stars (upper HD~199178, bottom AR~Lac) with the proper motion removed, showing position offset in right ascension (red line) and declination (blue line).}
 \label{pm_targets}
\end{figure}

\subsection{VLBA astrometric parameters}
\label{ssec:astrometry}

We used the \textsc{aips} task \texttt{JMFIT} to fit the brightness distribution in each 
phase-referenced image to a single elliptical Gaussian model. The position offsets of the stars
including their formal uncertainties relative to the background sources are summarised in Table~\ref{tab:targets_coords}. 

We fitted five astrometric parameters, i.e., parallax $\omega$, position offset (x, y) in the east-west direction and north-south direction at a reference epoch (at the middle of the first and last epochs), proper motions $\mu_{x}$ and $\mu_{y}$, to the position offsets at six-epoch for each source pair using the least square method. To evaluate the potential systematic errors mainly from the atmospheric delay, we added ``error floors'' in quadrature to the formal uncertainties in each coordinate, and adjusted them until the reduced chi-square of post-fit residual $\chi_\nu^2$ was close to unity.

The parallaxes of the two radio stars using different background sources are consistent with their uncertainties. 
We then combined the data for the two background sources to fit the five parameters in the same way as aforementioned, the fitted parameters and their uncertainties are listed in Table~\ref{tab:targets_fitted_results}.

\begin{table*}
	\centering
	\caption{Flux densities of background calibrators.}
	\label{calibs_fluxs}
	\small
	\begin{threeparttable}
	\begin{tabular}{cccccc}
	\hline
  	  Epoch   & Target  & Calibrator 1    & Flux density & Calibrator 2 & Flux density \\
      Code	  &         &                &  (Jy)         &         &  (Jy)        \\
	\hline      
        A  &  \multirow{6}{*}{HD199178} & \multirow{6}{*}{J2040+4527} & 0.0847$\pm$0.0042 & \multirow{6}{*}{J2102+4702} & 0.1379$\pm$0.0069\\
	B  &     &  & 0.0828$\pm$0.0041 &    & 0.1689$\pm$0.0084\\
	C  &     &  & 0.0758$\pm$0.0038 &    & 0.0981$\pm$0.0049\\
	D  &     &  & 0.0797$\pm$0.0040 &    & 0.0984$\pm$0.0049\\
	E  &     &  & 0.0864$\pm$0.0043 &    & 0.0966$\pm$0.0048\\
	F  &     &  & 0.0837$\pm$0.0042 &    & 0.0890$\pm$0.0045\\
        & & & & & \\
	A  &  \multirow{6}{*}{AR Lac}   & \multirow{6}{*}{J2153+4322} & 0.1936$\pm$0.0097 &\multirow{6}{*}{J2202+4216} & 1.9078$\pm$0.0954 \\
	B  &  &   & 0.1874$\pm$0.0094 &      & 1.8444$\pm$0.0922 \\
	C  &  &   & 0.1865$\pm$0.0093 &      & 2.0084$\pm$0.1004\\
	D  &  &   & 0.1968$\pm$0.0098 &      & 2.1873$\pm$0.1094\\
	E  &  &   & 0.1804$\pm$0.0090 &      & 3.5531$\pm$0.1777\\
	F  &  &   & 0.1807$\pm$0.0090 &      & 2.6227$\pm$0.1311\\
	\hline
	\end{tabular}
	\begin{tablenotes}   
        \footnotesize 
        \item \textbf{Note.} Total flux densities estimated from the \textsc{clean} components in images of the background calibrators. The quoted uncertainties are roughly estimated as 5\% of the total flux densities 
        \citep{2002ApJ...568...99H}. 
   \end{tablenotes}
	\end{threeparttable}
\end{table*}

\subsection{Flux densities of HD~199178 and AR~Lac}
\label{ssec:flux}

\begin{figure*}
	\centering
		\centering
        \begin{overpic}[width=0.32\linewidth]{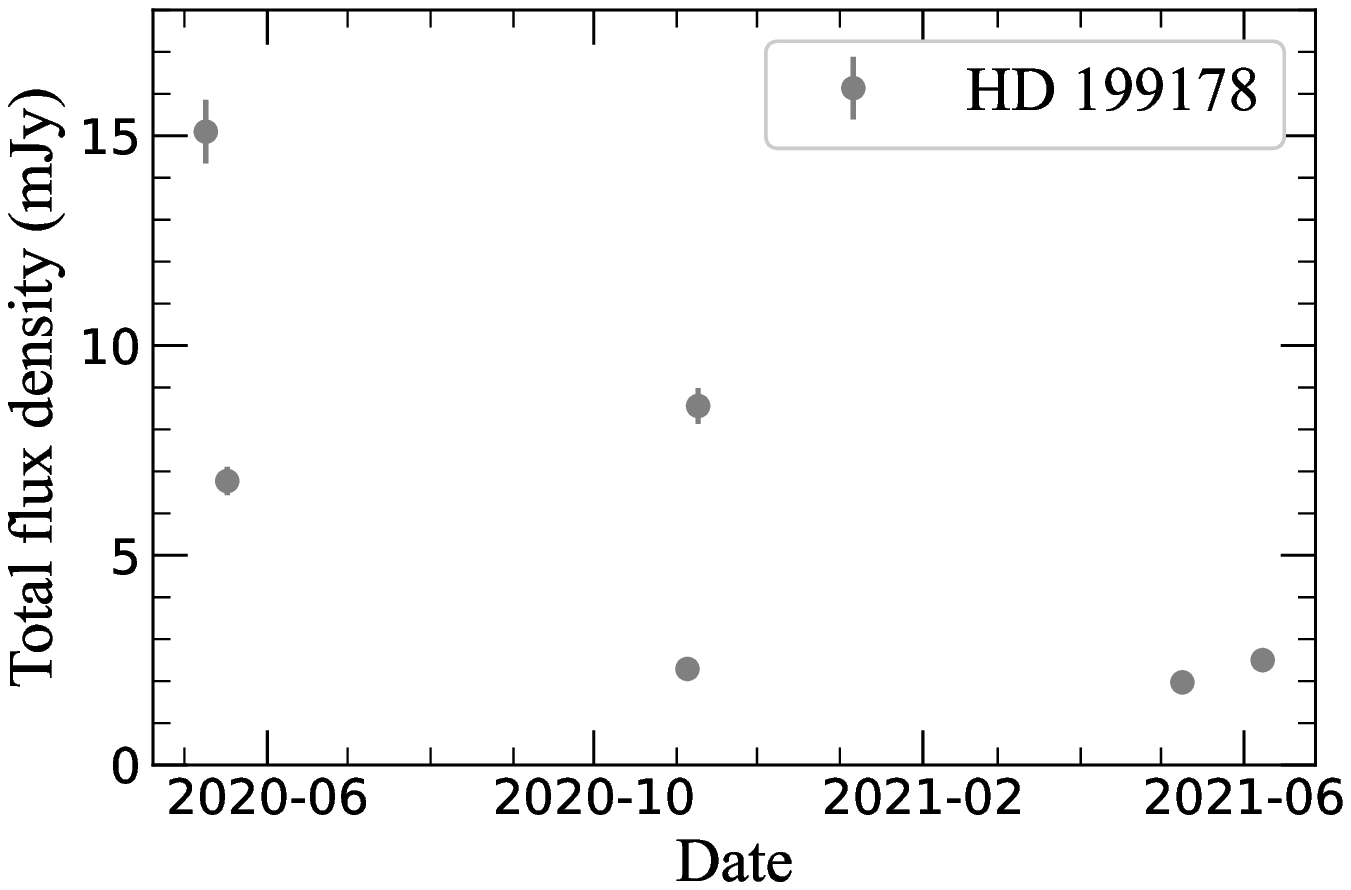}
        \end{overpic}
        \begin{overpic}[width=0.33\linewidth]{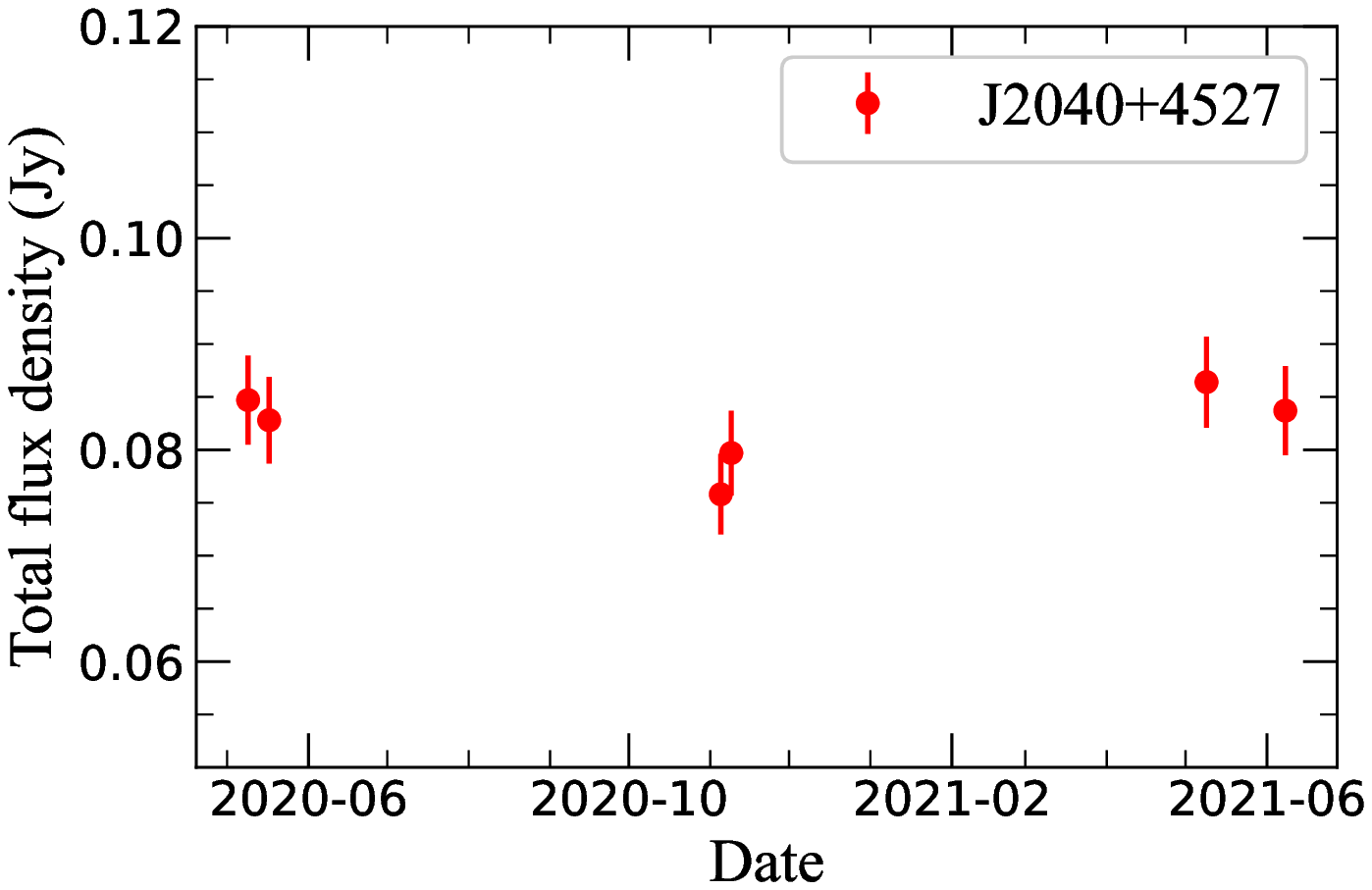}
        \end{overpic}
        \begin{overpic}[width=0.325\linewidth]{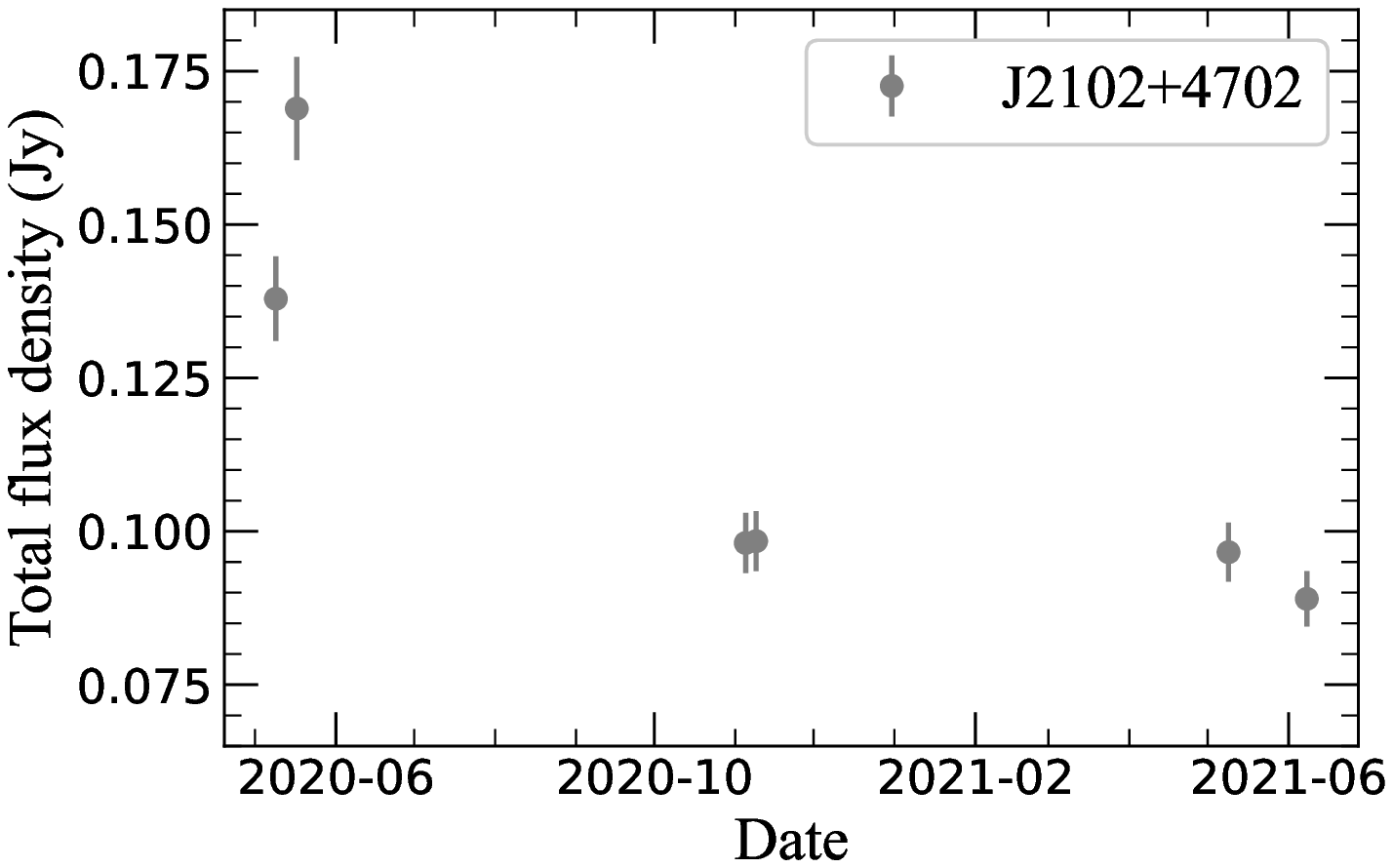}
        \end{overpic}
		\centering
		\begin{overpic}[width=0.325\linewidth]{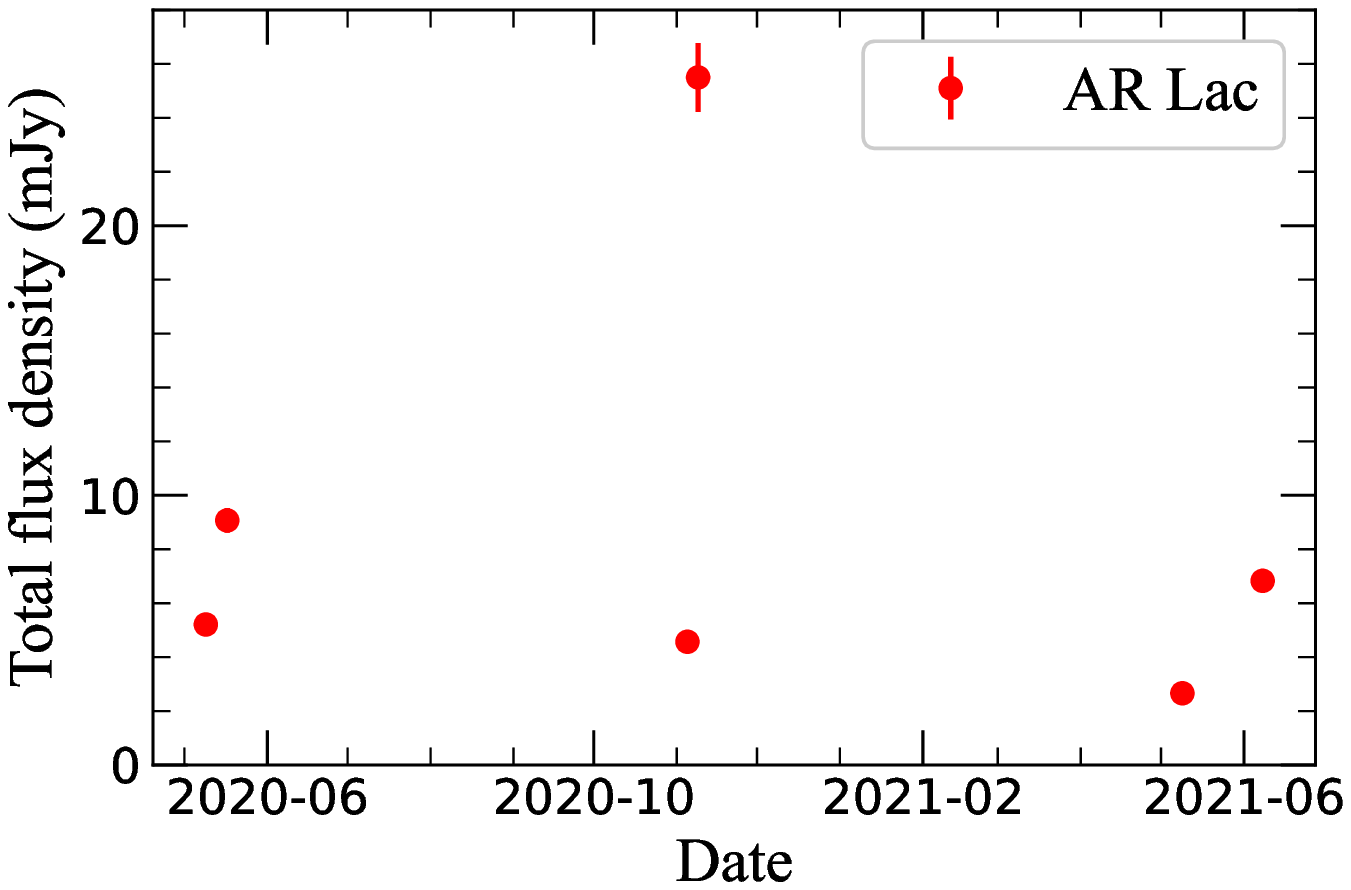}
        \end{overpic}
        \begin{overpic}[width=0.33\linewidth]{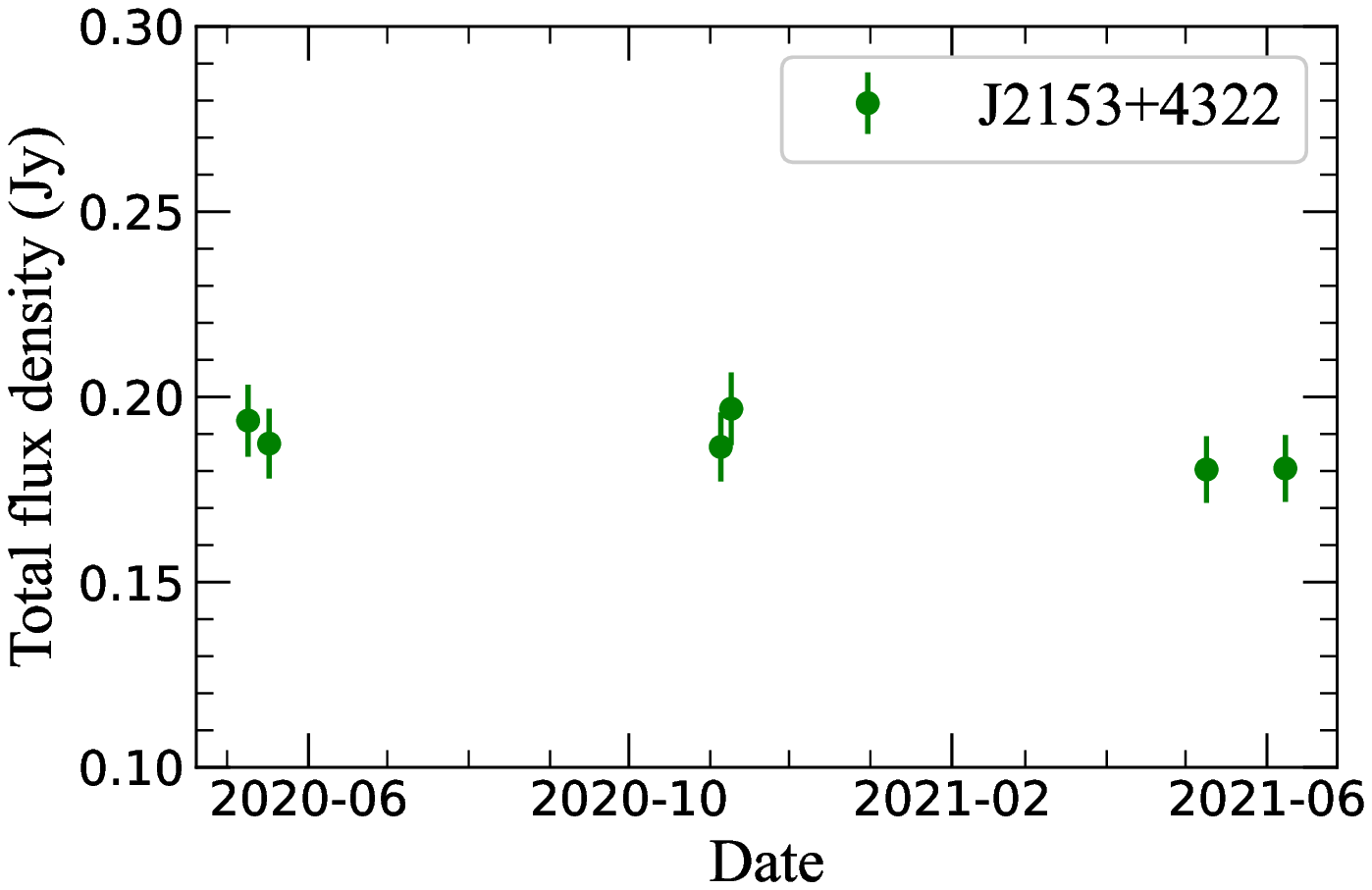}
        \end{overpic}
        \begin{overpic}[width=0.30\linewidth]{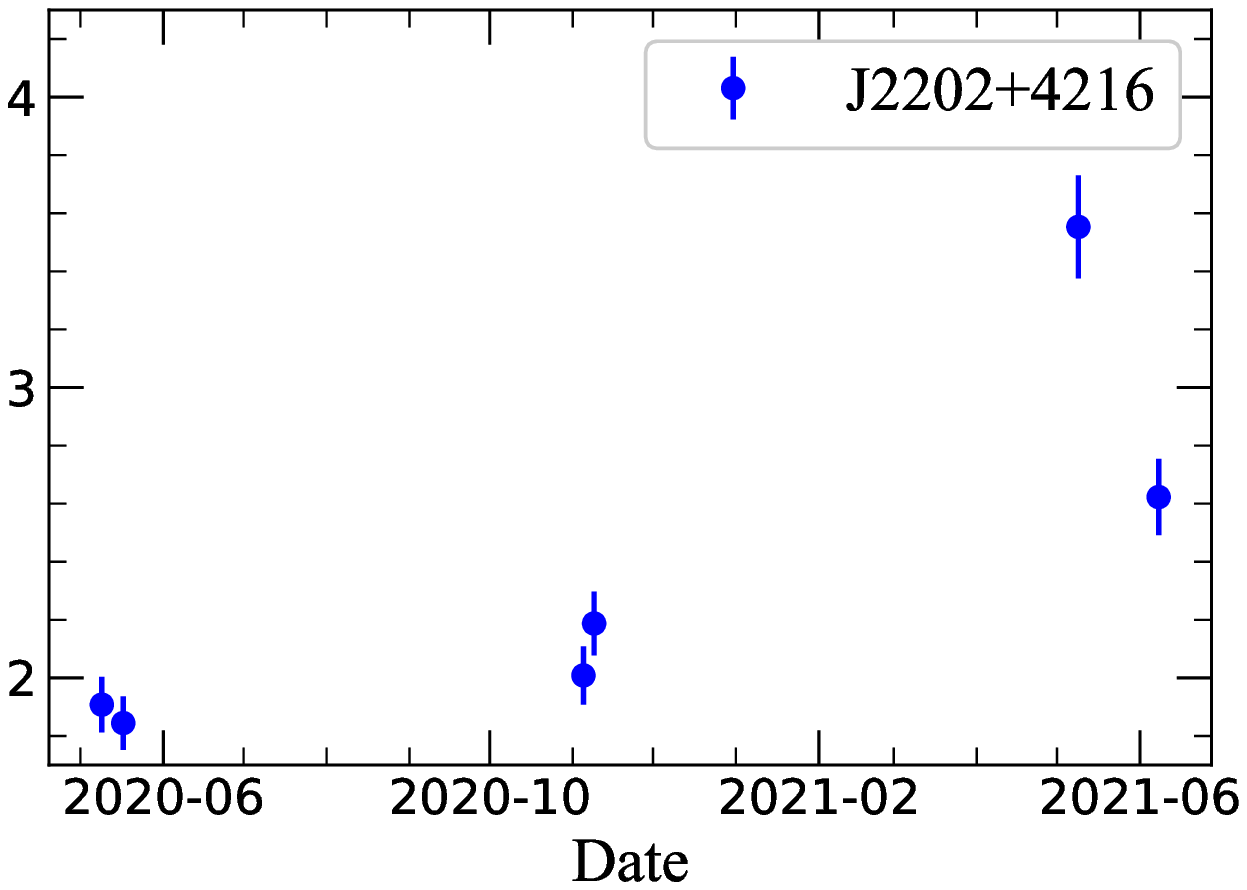}
        \end{overpic}		
\caption{Light curves of the two radio stars and their calibrators at 15 GHz. The flux densities of radio stars are obtained by combining two calibrators in Table~\ref{tab:targets_coords}. The uncertainties of radio stars included formal errors and 5\% systematic errors of the total flux densities, while the uncertainties of calibrators are only 5\% systematic errors.}
 \label{fig:flux_all}
\end{figure*}

The total flux densities of the radio stars of the calibrators are listed in Table~\ref{tab:targets_coords} and Table~\ref{calibs_fluxs}, respectively. The uncertainties of radio stars included formal errors and 5\% systematic errors \citep[empirical estimates,][]{2002ApJ...568...99H} of the total flux densities, while the uncertainties of calibrators are only 5\% systematic errors. As shown in Figure~\ref{fig:flux_all}, light curves of the radio stars and calibrators,
the two radio stars and the two calibrators J2202$+$4216 and J2102$+$4702, which are blazars
\citep{Bllac_variability1,Bllac_variability2},
show significant variations (the ratio of standard deviation to the mean value of the fluxes are greater than 61\%) on flux from epoch to epoch,  while the other two calibrators are relatively stable.      

The flux density of HD~199178 reached the maximum $\sim$16.09 mJy on 2020 May 9 and the minimum $\sim$1.80 mJy on 2021 May 9. Previous VLBI observations by 
\citet{1999A&A...344.1014L} and VLA + PT observations by \citet{VLA+PT_1} reported its flux density was a few mJy at 8.4 GHz, however, it had a large outburst of $\sim$300 mJy during the VLBI observations by
\citet{1999A&A...344.1014L}. 

For AR~Lac, the maximum and minimum flux densities during the observations were $\sim$25.58~mJy on 2020 November 9 and $\sim$2.73~mJy on 2021 May 9, respectively. \cite{1999A&A...344.1014L} reported the AR~Lac was detected all above 2 mJy at 8.4 GHz between April 1989 and May 1994 at 7 epochs from global VLBI array observations, and 
\cite{VLA+PT_1} obtained the flux density of 3.4 mJy at 8.4 GHz in December 2000 from the VLA + PT observations. 
In addition, relatively strong radio flare events had been discovered by earlier observations in the 1970s from the NRAO 3-element interferometer \citep{Arlac_flux_vari1} and VLA \citep{Arlac_flux_vari2}, the up to $\sim$10 times variability nature reached $\sim$120 mJy at 8.1 GHz and $\sim$28 mJy at 4.6 GHz within a day. This indicates flux of this radio star is highly variable and may relate to its orbit motion with a period of about 2 days.

In addition, these two radio stars showed clear flux density differences between right and left circular polarization. This indicates the existence of circular polarization, i.e., \textit{Stokes V}. Compared to the total flux densities \textit{Stokes I}, the fraction of $V/I$ varied from $\sim$10\% to $\sim$33\%. We have also checked the \textit{Stokes V} maps of the four calibrators, there are not any significant ($>5 \sigma$) features in the residual maps of \textit{Stokes V}. Therefore, the detection of circular polarization emission of the two radio stars, cannot result from the absence of $D$-term (signal leakage between the polarization channels) calibration~\citep{d-term_calib}.

The brightness temperature \textit{T$_{\rm b}$} can be estimated using the equation (\ref{eq1}) from \citep{1982ApJ...252..102C},

\begin{equation}
T_{\rm b} = 1.22\times10^{9}\frac{S_\mathrm{obs}}{\nu_\mathrm{obs}^2\theta_\mathrm{1}\theta_\mathrm{2}}(1+z),
\label{eq1}
\end{equation}
where $S_\mathrm{obs}$ is the observed total flux density in mJy, $\nu_\mathrm{obs}$ is the observing frequency in GHz, $\theta_\mathrm{1}$ and $\theta_\mathrm{2}$ are the FWHM of the Gaussian model in mas, and $z$ is the red-shift which is zero for Galactic radio stars. The estimated average brightness temperatures (T$_{b}$) of HD~199178 and AR~Lac are about
3.3$\times10^{7}$ K and 4.1$\times10^{7}$ K, respectively. This implies that the radio emission from
both stars are dominated by non-thermal radiation \citep{nonthermal1,nonthermal2}.

\section{Discussion}
\label{sec:disc}

\subsection{Intraday flux variation and polarization effects on parallax fits}
\label{ssec:pol}

As shown in Figure~\ref{HD_imgs} and \ref{ARLac_imgs},
images of both radio stars are not point-like.  
For HD 199178, it has an active magnetic strong emission from the corona and transition region~\citep{2019A&A...625A..79H},
while for AR Lac, the radio radiation is likely caused by the corona or collision of stellar winds of the two members of the close binary system~\citep{2001A&A...373..181T}.
This remind us that the intraday flux variation may mimic position shift and hence affect the parallax fit. To evaluate this effect, 
as shown in Figure~\ref{time_sched}, we divided the phase-referencing observations for each star at each epoch into three time slots
with an interval of about 160 minutes.
We have determined
the total flux densities and position offsets of the radio stars using the data from the three time slots,
then carried out the parallax and proper motion fits as described in Section~\ref{ssec:astrometry} separately. However, there are no significant differences in the estimated parallaxes and proper motions, i.e., the differences are within their joint uncertainties. This indicates that the motion of the radio emitting region dominated by the coronal mass ejection was small.

To investigate the polarization affect on position determination,  we made images from the left and
right circular polarization data, then determined the intensity and position offsets independently.
The ratio of circular polarization intensity (\textit{Stokes} $V$) and total intensity (\textit{Stokes} $I$) ranges
from $\sim$10\% to $\sim$30\% for both stars. However, the positions determined from the \textit{Stokes} $V$ and $I$ 
images show no clear discrepancies compared to their uncertainties.

\subsection{Orbital effects on parallax fits for AR~Lac}

AR Lac is an eclipsing close binary with known orbit period, 
we can estimate AR Lac's orbital phase using the equation (\ref{eq2}) adapted from \citep{2004AcA....54..207K},
\begin{equation}
HJD = 2452501.212  + E \times 1.983194
\label{eq2}
\end{equation}
Where $HJD$ is the Heliocentric Julian Date of the observation,
$E$ is the number of periods elapsed since the initial $JD(2452501.212)$ in the equation (\ref{eq2}), 
and the binary orbital period is 1.983194 days \citep{2004AcA....54..207K}. 
The orbital phase is the fractional part of $E$.

The orbit phases of the six epochs range from $\sim$0.32 to $\sim$0.57, the maximum phase change specific to each epoch will not exceed 0.14, which implies a maximum angular orbit motion no more than 
0.27~mas during the observation at each epoch. This is because the major axis of apparent orbit is 6.176$\times10^{6}$~km \citep{2000A&A...358.1007F}, corresponding to an orbital separation of about 0.97 $\pm$ 0.01 mas at a distance of 42.63 $\pm$ 0.17~pc from our VLBI measurement.
Since the apparent motion due to the parallax and proper motion effects in 10 hours are almost linear, after removing these effects, the orbit trajectory in the sky for an RS~Cvn type eclipsing binary system is almost a line segment, and it is possible to determine the position angle of the line segment (orbit plane) if the positions determined from the three time slots are accurate enough. However,
we failed to find the position angle.
As listed in Table~\ref{tab:targets_fitted_results} and \ref{tab:Coords_radiostars}, the position error of AR Lac is dominated by the
systematic error of $\approx$ 0.3 mas in both x and y directions, while the corresponding orbital movement was less than 0.27~mas, 
hence it is difficult to detect the orbit motion within the limited orbital phase range. However, since the orbit phases at the six epochs are within a
range of 0.14, the position offsets due to the orbit motions should be similar at the six epochs, thus there are only limited effects on
the parallax and proper motion fits.

\subsection{Comparison between previous and our VLBI measurements}

The two radio stars had been observed with VLBI~\citep{1999A&A...344.1014L} and connected 
radio interferometers (CRI), i.e., the Very Large Array plus one of the VLBA telescopes located in Pie Town (VLA+PT)~\citep{VLA+PT_1,VLA+PT_2} and the Multi-Element Radio Linked Interferometer Network (MERLIN)
\citep{merlin_astrometry}. Proper motions of
the two stars had been measured by VLA+PT, while only reliable proper motion of AR Lac
have been measured by MERLIN. 
We listed the astrometric parameters of the two stars from the aforementioned CRIs and our VLBA observations for
comparison. 
One can find that the uncertainties of the proper motions measured by the CRIs are much larger than those from VLBI measurements, 
this is mainly due to the much poorer angular resolution. The discrepancies in proper motion between CRI and VLBI are all smaller than the uncertainties of CRI.
For VLBI measurements, our VLBI parallax uncertainty is about four times smaller than that of \cite{1999A&A...344.1014L}, while the uncertainties of proper motions are 
at the same level. 
The VLBI observations by \citet{1999A&A...344.1014L} included sensitive telescopes like 
phased-VLA (with an effective diameter of 130~m), Effelsberg 100~m and the Deep Space Network 70~m at Goldstone. However, the observing frequencies were 5.0~GHz and 8.4~GHz,
and the bandwidth was 28 MHz at the data recording rate of 56 Mbps, which is much smaller than 2048 Mbps of our VLBA observations. 
This indicates that although there are large telescopes involved in the previous VLBI observations, the image sensitivity is comparable with that of a typical imaging noise 
of 0.1 mJy~beam$^{-1}$ for our VLBA observations with relatively smaller (25 m) telescopes. However, 
there were only roughly
atmospheric and ionospheric delay calibration with only one calibrator and the observing frequencies (5.0 and 8 GHz) were lower than 15~GHz of our observations,
this lead to larger position uncertainty at each epoch and hence larger parallax uncertainty for both stars. 
The proper motion uncertainties of previous and our recent VLBI observations are at the same level for AR~Lac, however, the observing time range of our observations is about 1 year while it was about 5 years in \cite{1999A&A...344.1014L}.

\subsection{Comparison between Gaia and our VLBI measurements}

\begin{figure}
 \centering
    \includegraphics[width=0.9\linewidth]{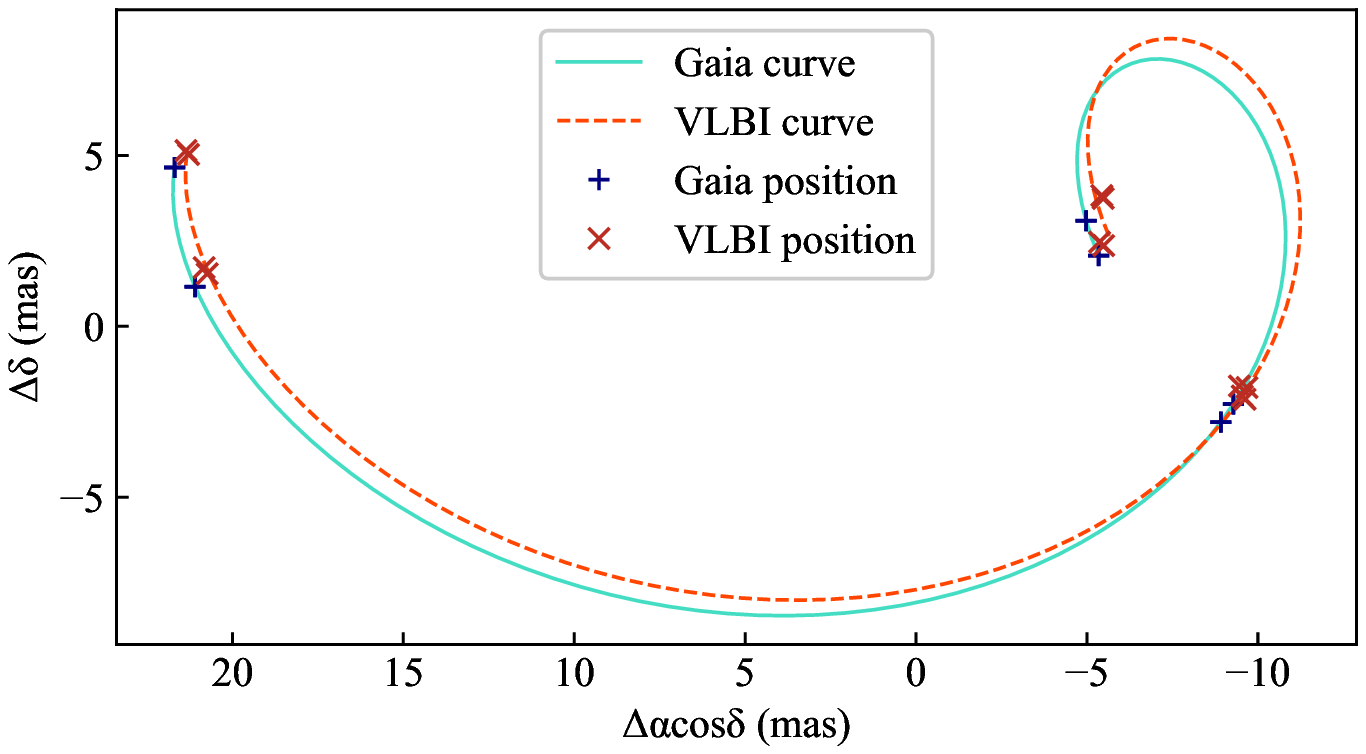}
    \includegraphics[width=0.9\linewidth]{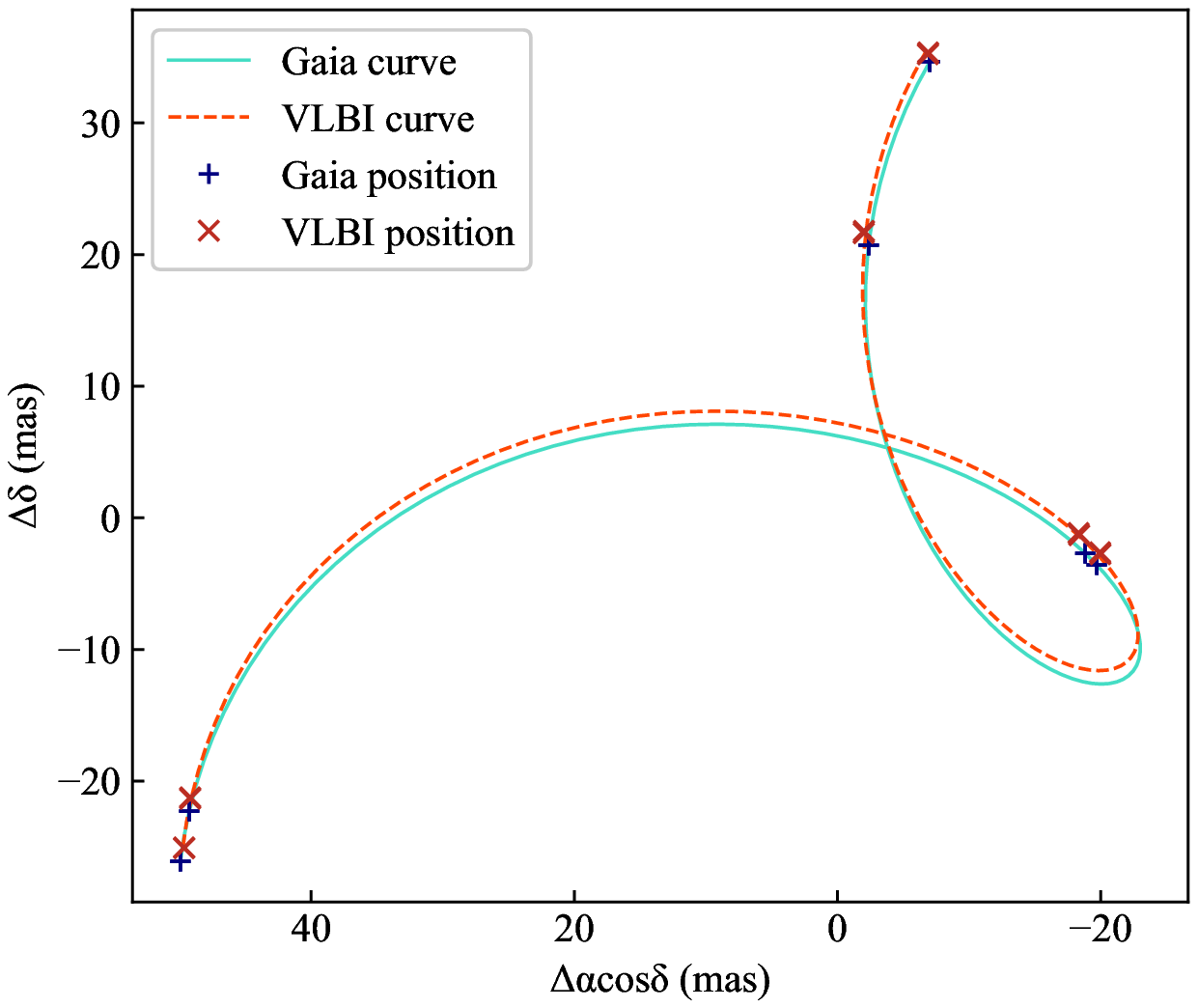}
    \caption{Positions comparison between \textit{Gaia} and our VLBI measurements.
Top panel: HD 199178; bottom panel: AR Lac.
The VLBI positions of each observation epoch are under ICRS and the offset values are relative to the reference positions ($\alpha$ = 20$^{\rm h}$53$^{\rm m}$53\fs 7034, $\delta$ = +44$^\circ$23$^\prime$11\farcs 0630 for HD~199178 and $\alpha$ = 22$^{\rm h}$08$^{\rm m}$40\fs 7140, $\delta$ = +45$^\circ$44$^\prime$33\farcs 0870 for AR~Lac).
The VLBI motion curves are drawn based on the parallax and proper motion estimated in this work.
The \textit{Gaia} positions and curves are based on \textit{Gaia} DR3 astrometric data, propagating to the VLBI observation epochs.}
\label{v_gaia}
\end{figure}

Fig.~\ref{v_gaia} shows the positions comparison between \textit{Gaia} and our VLBI measurements, systematic bias can be seen in both panels, which is likely because of the rotation between ICRF and GCRF, and other possible radio-optical offsets.

Our measured parallax ($\Pi$) of HD~199178 is $8.949 \pm 0.059$ mas corresponding to a distance of $111.70 \pm 0.70$~pc, while $\Pi$ of AR~Lac is $23.459 \pm 0.094$ mas corresponding to a distance of $42.63 \pm 0.17$~pc.
As shown in Table~\ref{tab:targets_fitted_results}, our VLBI measured astrometric parameters are well consistent with those from \textit{Gaia}~DR3 within their joint uncertainties. 
The uncertainties of our VLBI measurements of parallax and proper motion
are about 4 and 10 times of
those from \textit{Gaia} DR3, respectively. 
This is most likely owing to the much longer observing periods of \textit{Gaia}, especially for proper motion.

For the VLBI astrometry of the two radio stars, except for the target's orbit motion effects which are not taken into account in the current \textit{Gaia} DR3,
the primary error sources may include the following,

1) All the targets and calibrators are not point-like sources but show certain extended structures, especially for the calibrator J2202$+$4216 (see Fig.~\ref{calibs_img}) which has a clear southward jet structure. Although the self-calibrations on the calibrators have been carried out in data processing to reduce the uncertainties caused by the complex structures, the source structure effect cannot be completely eliminated.

2) For phase-referencing observation, the calibrators are not close enough to the target sources, i.e., with angular separation greater than 2\degr.
Although geodetic-like blocks were used to estimate the residual atmospheric delay which is usually the dominant error source at observing frequency $>$ 5~GHz, there are always uncompensated errors in the relative position measurements.

\subsection{Frame link with updated \textit{Gaia} and VLBI data}\label{sec:frame_link}

The connection parameters between GCRF and ICRF given by \citet{2020A&A...633A...1L, 2020A&A...637C...5L} are derived from 26 best-fitting radio stars, including both HD~199178 and AR~Lac.
For the orientation solution, stars with smaller astrometric parameter uncertainties (both VLBI and \textit{Gaia}) and proximity in time with \textit{Gaia} reference epoch will contribute more weight.
This is because propagating a star's astrometric data to a reference epoch will magnify its uncertainties in proportion to the epoch difference.
However, to the spin solution, a longer time span between VLBI and \textit{Gaia} epochs is better.
The astrometric data of these two stars were derived from 1990s' observations, which are $\sim$20 years before \textit{Gaia} DR3's reference epoch (J2016.0), resulting that HD~199178 and AR~Lac contributed a great amount of information on CRF spin parameters and very little information on orientation parameters in contrast.

Our new observations of the two stars are helpful supplements to the CRF connection.
Combining the astrometric data of the 26 radio stars from \textit{Gaia} DR3, VLBI data compiled by Lindegren
and our new data, we calculated the connection parameters between GCRF3 and ICRF3, and the result is shown in Table~\ref{crf_conn_res}.
Since only two stars' data is updated, the variation of the connection parameters is not significant.
Take solution C in Table~\ref{crf_conn_res} as an example, the data Lindegren used and our new data are both taken into calculation as separate sources.
Table~\ref{crf_conn_stat} gives the solution statistics for the stars, which shows that the new data of HD~199178 and AR~Lac we added contributes much more than the old data on orientation parameters (column $E_i$), while the contributions on spin parameters (column $\Omega_i$) are of the same level.

\begin{table*}
	\centering
	\caption{Contribution of the old and new data of HD~199178 and AR~Lac to the frame link}
	\label{crf_conn_res}
	\small
	\begin{threeparttable}
	\begin{tabular}{ccccccc}
    \hline
    Solution & \multicolumn{3}{c}{Orientation (mas) at $T={\rm J}2016.0$ } & \multicolumn{3}{c}{Spin (mas yr$^{-1}$)} \\
    & $\varepsilon_X(T)$ & $\varepsilon_Y(T)$ & $\varepsilon_Z(T)$ & $\omega_X$ & $\omega_Y$ & $\omega_Z$ \\
	\hline
    A & $ -0.006\pm0.102 $ & $ +1.196\pm0.476 $ & $ +0.621\pm0.079 $ & $ +0.010\pm0.031 $ & $ +0.129\pm0.038 $ & $ -0.004\pm0.035 $ \\
    B & $ -0.047\pm0.104 $ & $ +0.923\pm0.395 $ & $ +0.532\pm0.110 $ & $ -0.000\pm0.037 $ & $ +0.065\pm0.038 $ & $ -0.065\pm0.044 $ \\
    C & $ -0.049\pm0.096 $ & $ +0.915\pm0.359 $ & $ +0.535\pm0.101 $ & $ -0.001\pm0.030 $ & $ +0.075\pm0.029 $ & $ -0.056\pm0.038 $ \\
	\hline
	
	\end{tabular}
	\begin{tablenotes}
        \footnotesize
        \item \textbf{NOTE.} Solution A used 5-parameter astrometric data of 26 radio stars from VLBI compiled by \citet{2020A&A...633A...1L} and corresponding \textit{Gaia} DR3 data; on the basis of solution A, solution B replaced HD~199178's and AR~Lac's data with our new five-parameter result; solution C is the combination of Lindegren's and our new data, which treats the old and new data as separate stars.
        The three stars with two standalone observations used in Lindegren's dataset (HD~283572, Cyg~X-1, IM~Peg) are processed the same as HD~199178 and AR~Lac.
        The uncertainties of the parameters are estimated by the bootstrap resampling method.
        \end{tablenotes}
	\end{threeparttable}
\end{table*}

\begin{table}
	\centering
	\caption{}
	\label{crf_conn_stat}
	\small
	\begin{threeparttable}
	\begin{tabular}{ccccc}
	\hline
    Name & Epoch & $E_i$ & $\Omega_i$ & $Q_i/n_i$ \\
    & (Julian year) & (mas$^{-2}$) & (mas$^{-2}$yr$^2$) &  \\
	\hline
    HD 199178 & 2020.8939 & 250.5 & 5927.7 &8.032 \\
    HD 199178 & 1993.7933 & 8.2 & 4062.6 & 3.421 \\
    
    AR Lac    & 2020.8940 & 136.1 & 3239.8 & 8.386 \\
    AR Lac    & 1992.4353 & 6.2 & 3408.3 & 1.111 \\
	\hline
	\end{tabular}
	\begin{tablenotes}
        \footnotesize
        \item \textbf{NOTE.} The definitions of the symbols are kept the same as \citet{2020A&A...633A...1L}: $E_i$ and $\Omega_i$ are the formal weights contributed by the star to the estimation of $\varepsilon(T)$ and $\omega$, and $Q_i/n_i$ is the reduced chi-square of the star normalized by $n_i$, the number of VLBI data points included for the star.
   \end{tablenotes}
	\end{threeparttable}
\end{table}

\section{Summary}
\label{sec:sum}

Based on multiple epoch phase-referenced VLBA observations at 15~GHz, 
we have measured parallaxes and proper motions of two radio stars HD~199178 and AR~Lac,
with the primary goal to add new data of radio stars with improved astrometric parameters to link the radio and
optical CRFs. The parallax and proper motion uncertainties are at the level of tens of $\mu$as and $\sim$ 100 $\mu$as~yr$^{-1}$,
which are comparable with those uncertainties of \textit{Gaia}.  The frame link adding the new data of these two radio stars shows
a significant contribution to the orientation compared to the old data. 

This pilot program indicates not only the feasibility to 
link the radio and optical frames using the radio stars but also the possible improvement of orbit parameters, e.g., 
at least for the position angle of the major axis of the binary systems.

\section*{Acknowledgements}

We would like to thank the referee for very detailed and helpful comments. We acknowledge the NRAO VLBA network, NRAO is a facility of the NationalScience Foundation (NSF) operated under cooperative agreement by Associated Universities, Inc. This work was supported by the National Natural Science Foundation of China (NSFC) under grant Nos. U2031212 and 11903079.

\section*{Data Availability}

This paper makes use of VLBA data which codes are from BZ077A to BZ077E, available for download at \url{https://data.nrao.edu/portal/}.



\bibliographystyle{mnras}
\bibliography{ref} 








\bsp	
\label{lastpage}
\end{document}